\documentclass[aps,groupedaddress, superscriptaddress,showpacs,amsmath,amssymb,twocolumn,prl]{revtex4-1}
\usepackage{bm}
\usepackage[dvips]{graphicx}
\usepackage{wrapfig,subfigure}
\usepackage{amssymb}
\usepackage{color}
\usepackage[normalem]{ulem}
\usepackage{amsmath}
\usepackage{epstopdf}

\def\be{\begin{equation}}
\def\ee{\end{equation}}
\begin{document}
\title{Correlators in simultaneous measurement of non-commuting qubit observables}
\author{Juan Atalaya}
\affiliation{Department of Electrical and Computer Engineering, University of California, Riverside, CA 92521, USA}
\author{Shay Hacohen-Gourgy}
\affiliation{Quantum Nanoelectronics Laboratory, Department of Physics, University of California, Berkeley CA 94720, USA}
\affiliation{Center for Quantum Coherent Science, University of California, Berkeley CA 94720, USA.}
\author{Leigh S. Martin}
\affiliation{Quantum Nanoelectronics Laboratory, Department of Physics, University of California, Berkeley CA 94720, USA}
\affiliation{Center for Quantum Coherent Science, University of California, Berkeley CA 94720, USA.}
\author{Irfan Siddiqi}
\affiliation{Quantum Nanoelectronics Laboratory, Department of Physics, University of California, Berkeley CA 94720, USA}
\affiliation{Center for Quantum Coherent Science, University of California, Berkeley CA 94720, USA.}
\author{Alexander N. Korotkov}
\affiliation{Department of Electrical and Computer Engineering, University of California, Riverside, CA 92521, USA}
\date{\today}
\begin{abstract}
We consider the simultaneous and continuous measurement of qubit observables $\sigma_z$ and
$\sigma_z\cos\varphi + \sigma_x\sin\varphi$, focusing on the temporal correlations of the two output signals. Using quantum Bayesian theory, we derive analytical expressions for the correlators, which we find to be in very good agreement with experimentally measured output signals.
We further discuss how the correlators can be applied to parameter estimation, and use them to infer a small residual qubit Hamiltonian arising from calibration inaccuracy in the experimental data.
\end{abstract}

\pacs{}
\maketitle

Continuous quantum measurements (CQMs) have become a unique platform to explore fundamental aspects of quantum phenomena and have potential applications to quantum information science.
They have been discussed theoretically
in various contexts
(e.g., \cite{Kraus1983, Caves1986, Menskii1998, Belavkin1992, Braginsky-book, Aharonov1998, Dalibard1992, Carmichael1993, Wiseman1993, Korotkov1999}), and in the past decade superconducting qubits have become the main experimental system for the realization of CQMs \cite{Katz2006, Laloy2010, Vijay2012, Hatridge2013, Murch2013, deLange2014, Campagne2014}. CQMs are shedding new light on our understanding of the quantum measurement process, and there is also a growing interest in CQM applications, including quantum feedback \cite{Wiseman1993fb, Ruskov2002, Sayrin2011, Vijay2012, deLange2014}, rapid state purification \cite{Jacobs2003}, preparation of entangled states \cite{Ruskov2003,Riste2013,Roch2015}, and continuous quantum error correction \cite{Ahn2002,Sarovar2004}.

While a simultaneous measurement of non-commuting observables is impossible with projective measurements, nothing theoretically forbids such a measurement using CQMs. Aside from new physics, such a protocol may open up new areas of applications, inaccessible to projective measurements.
The theoretical discussion of a simultaneous measurement of incompatible observables has a long history \cite{Arthurs1965, Busch1985, Stenholm1992, Jordan2005}. For the measurement of non-commuting observables of a qubit, statistics of time-integrated detector outputs and fidelity of state monitoring directly via time-integrated outputs has been analyzed in Ref.\ \cite{Wei2008}. The evolution of the qubit state due to simultaneous measurement of incompatible variables has been described theoretically in Ref.\ \cite{Ruskov2010}, and has been recently demonstrated experimentally in Ref.\ \cite{Shay2016}.

\begin{figure}[tb]
\centering
\includegraphics[width=6.5cm, trim = 2.0cm 2.0cm 0.0cm 1.5cm,clip=true]{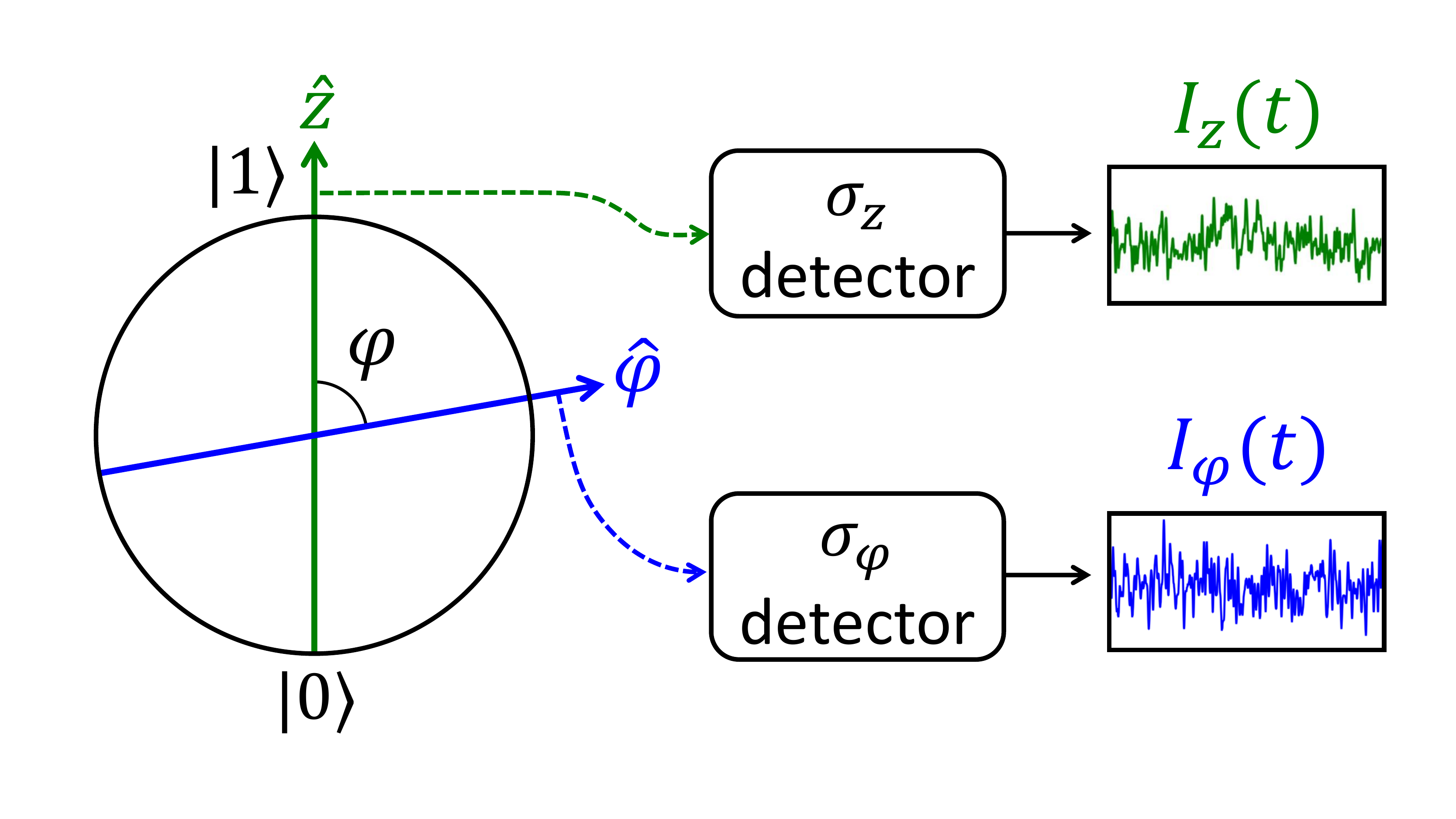}
\caption{We consider the simultaneous continuous measurement of qubit observables $\sigma_z$ and $\sigma_\varphi$, which differ by an angle $\varphi$ on the Bloch sphere, and calculate time-correlators for the output signals $I_z(t)$ and $I_\varphi(t)$ resulting from this measurement.}
\label{fig:schematic}
\end{figure}

In this letter, we focus on the temporal correlations of the output signals from two linear detectors measuring continuously and simultaneously the qubit observables $\sigma_z$ and $\sigma_\varphi\equiv \sigma_z\cos\varphi + \sigma_x\sin\varphi$, where $\sigma_x$ and $\sigma_z$ are the Pauli matrices and $\varphi$ is an angle between the two measurement directions on the Bloch sphere (Fig.\ \ref{fig:schematic}). The experimental setup is described in detail in Ref.~\cite{Shay2016}; it is based on a Rabi-rotated physical qubit, which is measured stroboscopically \cite{Averin2002} using symmetric sideband pumping of a coupled resonator, so that $\sigma_z$ and $\sigma_\varphi$ for an effective rotating-frame qubit are being measured. Description of such a measurement based on the theory of quantum trajectories \cite{Carmichael1993, Wiseman1993, Wiseman-book, Gambetta2008} has been developed in Ref.\ \cite{Shay2016}. In this letter we will use a simpler approach based on quantum Bayesian theory \cite{Korotkov1999,Korotkov2001-16}. The quantum Bayesian description of the rotating-frame experiment \cite{Shay2016} is developed in the Supplemental Material \cite{supplement}. The goal of this letter is calculation of the time-correlators for the output signals measuring $\sigma_z$ and $\sigma_\varphi$, and their comparison with experimental data.
As we will see, these correlators may be a useful tool for sensitive parameter estimation in an experiment. These correlators are also important in the analysis of quantum error detection and correction based on simultaneous measurement of non-commuting operators \cite{Atalaya2016}. We note that the analyzed output signal correlators are different from qubit-state correlators \cite{Chantasri2015}.

{\it Quantum Bayesian theory.}---A simultaneous continuous measurement of the qubit observables $\sigma_z$ and $\sigma_\varphi$ by two linear detectors (Fig.\ \ref{fig:schematic}) produces noisy output signals $I_z(t)$ and $I_\varphi(t)$, respectively \cite{Ruskov2010, Korotkov2001-16},
\begin{align}
\label{eq:outputs-Iz}
I_z(t) =&\, {\rm Tr}[\sigma_z\rho (t)] + \sqrt{\tau_z}\, \xi_z(t), \\
I_\varphi(t) =&\, {\rm Tr}[\sigma_\varphi\rho (t)] + \sqrt{\tau_\varphi}\, \xi_\varphi(t),
\label{eq:outputs-Iphi}
\end{align}
where $\rho (t)$ is the qubit density matrix and $\tau_z$, $\tau_\varphi$ are the ``measurement'' (collapse) times needed for an informational signal-to-noise ratio of 1 for each channel. Note the chosen normalization for $I_z$ and $I_\varphi$.
In the Markovian approximation, the noises $\xi_z$ and $\xi_\varphi$ are uncorrelated, white, and Gaussian with two-time correlators
    \begin{align}
\label{eq:xi-corr}
\langle\xi_{z}(t)\, \xi_{z}(t')\rangle = \langle\xi_{\varphi}(t)\, \xi_{\varphi}(t')\rangle = \delta(t-t')
    \end{align}
and $\langle\xi_{z}(t)\, \xi_{\varphi}(t')\rangle = 0$. The qubit state is characterized in the Bloch-sphere representation as $\rho (t)\equiv \left[\openone + x(t)\, \sigma_x + y(t)\, \sigma_y + z(t)\, \sigma_z\right]/2$. We assume phase-sensitive amplifiers in the experimental setup, amplifying the optimal (informational) quadratures, so that the qubit evolution due to measurement is not affected by the phase backaction related to fluctuations in the orthogonal (non-informational) quadrature \cite{Wiseman-book, Gambetta2008, Korotkov2001-16}. Then there is only the quantum informational backaction, which for measurement of $\sigma_z$ and $\sigma_\varphi$ is described \cite{Ruskov2010,Korotkov2001-16,supplement} by the evolution equations (in the It\^o interpretation)
    \begin{align}
\dot x=&\, - \Gamma_z x - \Gamma_{\varphi}  \cos\varphi\left(x\cos\varphi - z\sin\varphi\right)  - \tau_z^{-1/2}xz \, \xi_z  \nonumber \\
          &\,  - \tau_\varphi^{-1/2}\left[xz\cos\varphi - (1-x^2)\sin\varphi\right] \xi_\varphi , \label{eq:Ito-x} \\
\dot y=&\, - (\Gamma_z +\Gamma_\varphi )\, y - \tau_z^{-1/2}yz \, \xi_z
           \nonumber \\
          &\,  - \tau_\varphi^{-1/2} y \left[z\cos\varphi + x\sin\varphi\right] \xi_\varphi , \label{eq:Ito-y} \\
\dot z=&\, \Gamma_{\varphi}  \sin\varphi \left(x\cos\varphi - z\sin\varphi\right) +
              \tau_z^{-1/2}(1-z^2)\, \xi_z    \nonumber \\
          &\, + \tau^{-1/2}_\varphi\left[(1 - z^2)\cos\varphi -xz\sin\varphi\right]\xi_\varphi .
\label{eq:Ito-z}
\end{align}
Here $\Gamma_z$ and $\Gamma_\varphi$ are the ensemble dephasing rates due to measurement, so that the quantum efficiencies \cite{Korotkov2001-16} for the two channels are
$\eta_{z} = 1/(2\tau_z\Gamma_z)$ and $\eta_\varphi = 1/(2\tau_\varphi\Gamma_\varphi)$. In the experiment $\eta_z =0.49$ and $\eta_\varphi=0.41$.

Equations \eqref{eq:Ito-x}--\eqref{eq:Ito-z} describe qubit evolution only due to measurement. We also need to add terms due to unitary evolution and due to decoherence not related to measurement. We assume the Hamiltonian $H=\hbar \tilde\Omega_{\rm R} \sigma_y/2$, describing Rabi oscillations about $y$-axis with frequency $\tilde\Omega_{\rm R}$. In the experiment, $\tilde\Omega_{\rm R}=\Omega_{\rm R}-\Omega_{\rm rf}$ is a small (kHz-range) undesired mismatch between the physical qubit Rabi frequency $\Omega_{\rm R}$ and rotating frame frequency $\Omega_{\rm rf}$ defined by detuning of sideband pumps \cite{Shay2016, supplement}. Decoherence of the effective qubit arises from the decoherence of the physical qubit with energy relaxation time $T_1$ and dephasing time $T_2$ [the pure dephasing rate is then $T_{\rm pd}^{-1}=T_2^{-1}-(2T_1)^{-1}$]. Averaging the decoherence over fast rotation $\Omega_{\rm R}\gg T_{2}^{-1}$ and adding unitary evolution, we obtain \cite{supplement}
\begin{align}
& \dot x = \tilde\Omega_{\rm R} z - \gamma x,\;\;\; \dot y = -T_2^{-1} y, \;\;\; \dot z = -\tilde\Omega_{\rm R} x -\gamma z,
 \label{eq:extra-evol} \\
& \gamma = (T_1^{-1}+T_{2}^{-1})/2, \,\,\,\,
\label{eq:gamma}\end{align}
Evolution of the effective qubit is described by adding terms from Eqs.\ \eqref{eq:Ito-x}--\eqref{eq:Ito-z} and (\ref{eq:extra-evol}).

{\it Correlators.}---Our next goal is to calculate the two-time correlators, $K_{ij}(\tau)$, for the output signals,
    \begin{align}
\label{eq:corr-def}
K_{ij}(\tau) \equiv  \langle I_j(t_1+\tau)\, I_i(t_1)\rangle,\,\,\,\,\, \tau>0, \,\,\,\,\, i,j \in \{z,\varphi\}.
    \end{align}
Self- and cross-correlators correspond to $i=j$ and $i\neq j$, respectively. The averaging in Eq.\ (\ref{eq:corr-def}) is over an ensemble of measurements with the initial qubit state $\rho _{\rm in}$ prepared at time $t_{\rm in}\leq t_1$. We will see, however, that  the result does not depend on $\rho_{\rm in}$, $t_{\rm in}$, and $t_1$, so Eq.\ (\ref{eq:corr-def}) can also be understood as averaging over time $t_1$. We assume that the parameters in Eqs.\ \eqref{eq:Ito-x}--(\ref{eq:extra-evol}) do not change with time. By assuming $\tau>0$, we avoid considering the trivial zero-time contribution to the self-correlators, $\Delta K_{ii}(\tau )=\tau_i\, \delta (\tau)$.

As shown in the Supplemental Material \cite{supplement}, calculation of the correlators from Eqs.\ (\ref{eq:outputs-Iz})--(\ref{eq:extra-evol}) is equivalent to the following {\it recipe} \cite{Korotkov2001sp}: we replace an actual continuous measurement at the (earlier) time moment $t_1$ with a projective measurement of $\sigma_i$, so that the measurement result $I_i(t_1)$ is $\pm 1$ with probability $\{1\pm {\rm Tr}[\sigma_i\, \rho(t_1)]\}/2 $, and the qubit state collapses correspondingly to the eigenstate $|1_i\rangle$ or $|0_i\rangle$ of $\sigma_i$ ($\sigma_i|1_i\rangle =|1_i\rangle$, $\sigma_i|0_i\rangle =-|0_i\rangle$). This gives the correlator
    \begin{align}
& K_{ij}(\tau) = {\rm Tr}[\sigma_j \, \rho_{\rm av}(t_1+\tau|1_i)]\, \frac{1+ {\rm Tr}[\sigma_i\, \rho(t_1)]}{2}
    \nonumber \\
 &\hspace{1.3cm}-  {\rm Tr}[\sigma_j \, \rho_{\rm av}(t_1+\tau|0_i)] \, \frac{1- {\rm Tr}[\sigma_i \,\rho(t_1)]}{2} ,
    \end{align}
where $\rho_{\rm av}(t_1+\tau|1_i)$ is the ensemble-averaged density matrix at time $t_1+\tau$ with the initial condition $\rho_{\rm av}(t_1|1_i)=|1_i\rangle\langle 1_i|$; similarly,  $\rho_{\rm av}(t_1+\tau|0_i)$ starts with $\rho_{\rm av}(t_1|0_i)=|0_i\rangle\langle 0_i|$. The evolution of $\rho_{\rm av}$ is given by Eqs.\ (\ref{eq:Ito-x})--(\ref{eq:extra-evol}) without noise, $\xi_z=\xi_\varphi=0$ (because of the It\^o form), so that
 \begin{align}
\dot x_{\rm av}=&\, - \Gamma_z x_{\rm av} - \Gamma_{\varphi}  \cos\varphi\left( x_{\rm av}\cos\varphi - z_{\rm av}\sin\varphi\right)
    \nonumber \\
 & +\tilde\Omega_{\rm R} z_{\rm av} -\gamma x_{\rm av},
    \label{eq:Ito-x-aver} \\
\dot y_{\rm av}=&\, - (\Gamma_z +\Gamma_\varphi )\, y_{\rm av} -T_2^{-1} y_{\rm av}  ,
    \label{eq:Ito-y-aver} \\
\dot z_{\rm av}=&\, \Gamma_{\varphi}  \sin\varphi \left( x_{\rm av}\cos\varphi - z_{\rm av}\sin\varphi\right) - \tilde\Omega_{\rm R} x_{\rm av} -\gamma z_{\rm av} .
\label{eq:Ito-z-aver}
\end{align}
These equations have an analytical solution presented in \cite{supplement} (note that the evolution of the $y$-coordinate is not important in our analysis). Thus we obtain the following correlators (alternative methods for the derivation are also discussed in \cite{supplement}):
    \begin{align}
\label{eq:K-zz}
 & K_{zz}(\tau) = \frac{1}{2} \bigg[ 1+\frac{\Gamma_z+\cos(2\varphi)\,\Gamma_\varphi}{\Gamma_+ -\Gamma_-}\bigg]e^{-\Gamma_-\tau }
    \nonumber \\
& \hspace{1.2cm}  + \frac{1}{2} \bigg[1 - \frac{\Gamma_z+\cos(2\varphi)\,\Gamma_\varphi}{\Gamma_+ -\Gamma_-}\bigg]\,e^{-\Gamma_+\tau },
    \\ \label{eq:K-zphi}
 & K_{z\varphi}(\tau) = \frac{(\Gamma_z + \Gamma_\varphi) \cos\varphi  + 2\tilde\Omega_{\rm R}\sin\varphi }{2(\Gamma_+ - \Gamma_-)}  \left(e^{-\Gamma_-\tau} -
e^{-\Gamma_+\tau}\right)
    \nonumber \\
&\hspace{1.2cm} +\frac{\cos \varphi}{2}\left( e^{-\Gamma_-\tau} + e^{-\Gamma_+\tau}\right),
    \\
& \hspace{0cm} \Gamma_{\pm} =\frac{\Gamma_z +\Gamma_\varphi \pm \big[ \Gamma_z^2 + \Gamma_\varphi^2 +2\Gamma_z\Gamma_\varphi\cos(2\varphi) -4 \tilde\Omega_{\rm R}^2\big]^{1/2} }{2}
    \nonumber \\
&\hspace{0.9cm}  +  (T_1^{-1}+T_2^{-1})/2.
\label{eq:Gamma_pm}
    \end{align}
Because of the rotational symmetry, the results for the correlators $K_{\varphi\varphi}(\tau)$ and $K_{\varphi z}(\tau)$ can be obtained from Eqs.\ \eqref{eq:K-zz} and \eqref{eq:K-zphi} by exchanging $\Gamma_z\leftrightarrow\Gamma_\varphi$ and $\varphi\to -\varphi$. The rotational symmetry also makes the correlators insensitive to a $y$-rotation in both measurement directions, $z\rightarrow \varphi_{\rm add}$, $\varphi\rightarrow \varphi+\varphi_{\rm add}$, by any angle $\varphi_{\rm add}$.

We emphasize that the obtained correlators do not depend on the qubit state $\rho (t_1)$ and therefore on $\rho_{\rm in}$ and $t_{\rm in}$ (this property would not hold in the presence of phase backaction). We also emphasize that the correlators depend on $\Gamma_{z}$ and $\Gamma_{\varphi}$, but do not depend on $\tau_{z}$ and $\tau_{\varphi}$ and therefore on the quantum efficiencies $\eta_z$ and $\eta_\varphi$. Physically, this is because non-ideal detectors can be thought of as ideal detectors with extra noise at the output \cite{Korotkov2001-16}, which only affects the zero-time self-correlators $K_{ii}(0)$.

Let us discuss some special cases for the results (\ref{eq:K-zz})--(\ref{eq:Gamma_pm}). (i) At small times, $\tau\to +0$, we obtain correlators
\begin{align}
\label{eq:small_time_limit}
K_{zz}(+0)= 1,\;\;\; K_{z\varphi}(0) = K_{\varphi z} (0)= \cos\varphi.
\end{align}
(ii) For $|\varphi|\ll 1$ and sufficiently small $T_2^{-1}$ and $\tilde\Omega_{\rm R}$, we have Zeno pinning near the states $|0\rangle$ and $|1\rangle$ with rare jumps between them with equal rates $\Gamma_{\rm jump}$. This produces cross-correlator \cite{Korotkov2011} $K_{z\varphi}(\tau)\approx \exp(-2\Gamma_{\rm jump}\tau)$ with jump rates
    \be
\Gamma_{\rm jump} = \frac{\varphi^2\Gamma_z\Gamma_\varphi + \tilde\Omega_{\rm R}^2} {2(\Gamma_z+\Gamma_\varphi )}  +  (T_1^{-1}+T_2^{-1})/4 .
    \ee
(iii) In the case $\tilde\Omega_{\rm R}=T_1^{-1}=T_2^{-1}=0$, we have full correlation for $\varphi=0$,  $K_{z\varphi}(\tau) = K_{zz}(\tau ) = 1$, full anticorrelation for $\varphi=\pi$, $K_{z\varphi}(\tau) = -K_{zz}(\tau ) = -1$, and no correlation for $\varphi=\pi/2$, $K_{z\varphi}(\tau) = 0$, while $K_{zz}(\tau )=e^{-\Gamma_\varphi \tau}$ and $K_{\varphi\varphi}(\tau )=e^{-\Gamma_z \tau}$. (iv) In the case $\tilde\Omega_{\rm R}=0$, the cross-correlator is symmetric, $K_{z\varphi}(\tau )=K_{\varphi z}(\tau)$, for any $\varphi$.

\begin{figure}[t!]
\centering
\begin{tabular}{cc}
\includegraphics[width=\linewidth, trim = 0cm 0cm 0cm
0.5cm,clip=true]{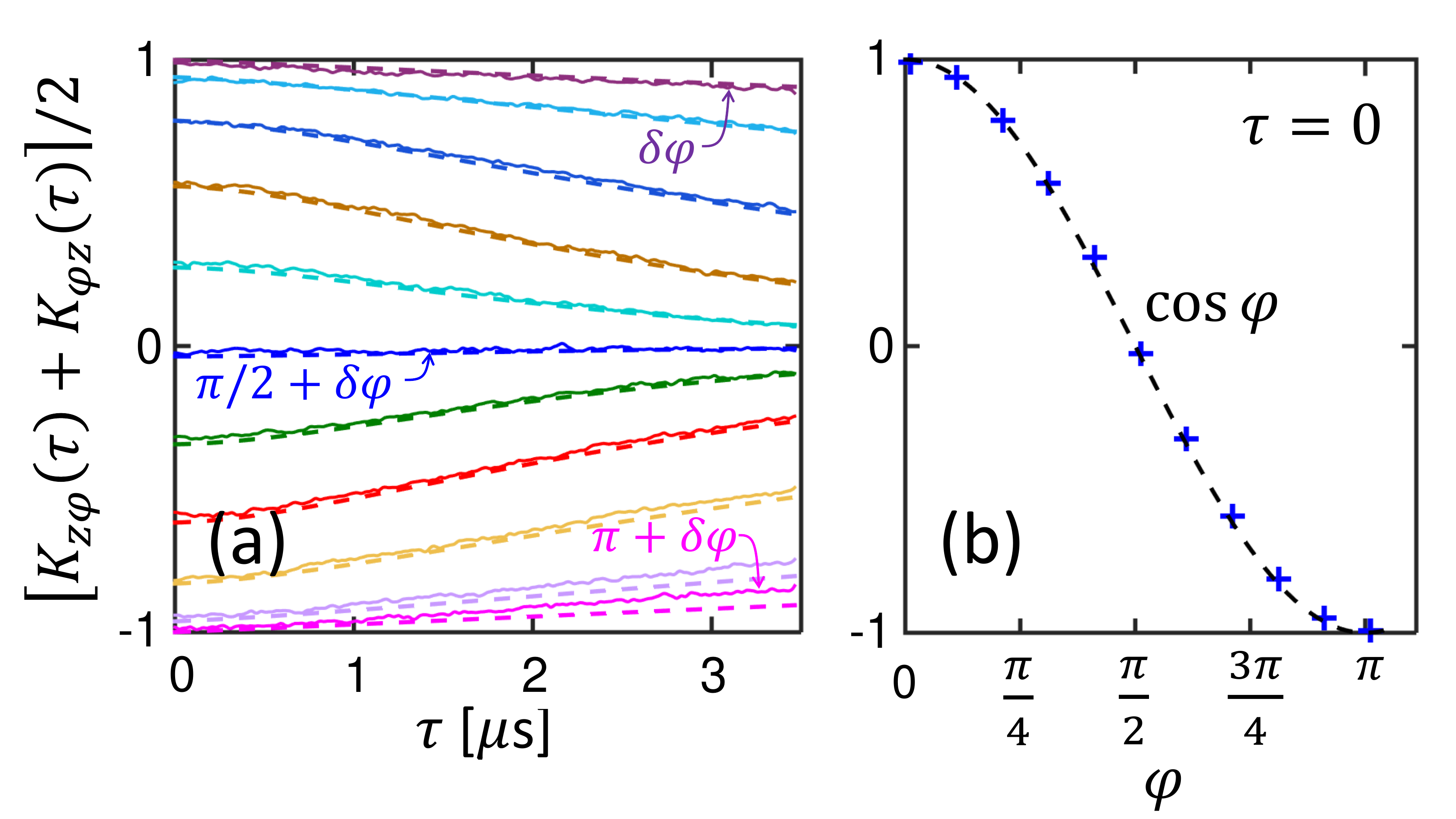}
\\
\includegraphics[width=\linewidth, trim=0cm 0cm 0cm
1.5cm,clip=true]{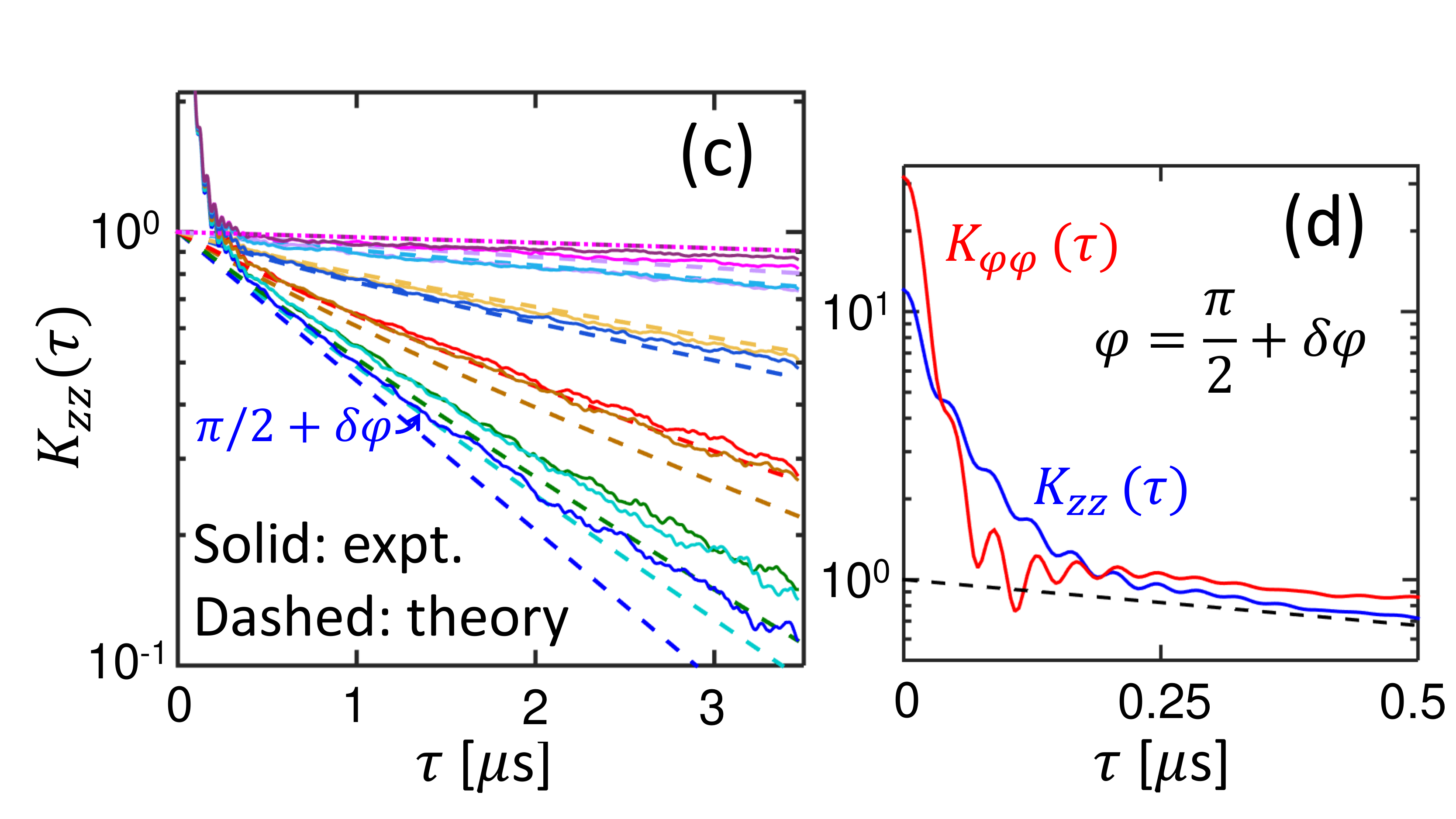}
\end{tabular}
\caption{Comparison between normalized experimental and theoretical correlators for the detector output signals.
We used 11 angles between the measurement axes: $\varphi = \varphi_n +\delta \varphi$ where $\varphi_n=n\pi/10$, $n=0, 1, \dots 10$ and $\delta\varphi \approx 0.036$. Solid and dashed lines in all panels correspond to experimental and analytical results, respectively.
Panel (a) shows the symmetrized cross-correlator $[K_{z\varphi}(\tau)+K_{\varphi z}(\tau)]/2$ for 11 values of $\varphi$, from $n=0$ (top) to $n=11$ (bottom). In panel (b) the crosses show $\varphi$-dependence of experimental cross-correlators from the panel (a) at $\tau =0$, while the dashed line, $\cos \varphi$, corresponds to Eq.\ \eqref{eq:small_time_limit}.
Panel (c) shows the self-correlator $K_{zz}(\tau)$ for 11 values of $\varphi$ [$n=0$ and 10 at the top, $n=5$ at the bottom, the same colors as in (a)]. Panel (d) illustrates deviation of experimental self-correlators (for $\varphi_n=\pi/2$) from the theory at small $\tau$ due to finite bandwidth of amplifiers and filters.
   }
\label{fig:theory-vs-exp}
\end{figure}

{\it Comparison with experimental results.}---Experimental data have been taken in the same way as in Ref.\ \cite{Shay2016} (see also \cite{supplement}). Experimental  parameters correspond to well-separated frequency scales, as needed for the theoretical results, $(T_1^{-1},T_2^{-1},|\tilde \Omega_{\rm R}|)\ll (\Gamma_z, \Gamma_\varphi) \ll (\kappa_{z},\kappa_{\varphi})\ll \Omega_{\rm R}$, with $T_1=60 \, \mu$s, $T_2=30\, \mu$s, $\Gamma_z^{-1}=\Gamma_\varphi^{-1}= 1.3\, \mu{\rm s}$, damping rates of the two measurement resonator modes  $\kappa_z/2\pi=4.3$ MHz and $\kappa_\varphi/2\pi=7.2$ MHz, and $\Omega_{\rm R}\approx \Omega_{\rm rf}= 2\pi \times 40$ MHz.
For this work we use 11 values for the angle $\varphi$ between the Bloch-sphere directions of simultaneously measured qubit observables: $\varphi_n=n\pi/10$, with integer $n$ between 0 and 10. While $\varphi_n$ is determined by well-controlled phases of applied microwaves \cite{Shay2016}, the effective $\varphi$ includes a small correction $\delta\varphi =(\kappa_\varphi -\kappa_z)/2\Omega_{\rm R} \approx 0.036$  (see \cite{supplement}), so that $\varphi =\varphi_n+\delta\varphi$. We have used about 200,000 traces per angle for the output signals $\tilde I_z(t)$ and $\tilde I_\varphi (t)$, each with 5 $\mu$s duration and 4 ns sampling interval.
The traces are selected by heralding the ground state of the qubit at the start of a run and checking that the transmon qubit is still within the two-level subspace after the run. The recorded signals $\tilde{I}_i(t)$ are linearly related to the normalized signals $I_i(t)$ in Eqs.\ (\ref{eq:outputs-Iz}) and (\ref{eq:outputs-Iphi}) as  $\tilde{I}_i(t)=(\Delta \tilde{I}_i/2)\, I_i(t)+\tilde{I}_i^{\rm off}$, where responses $\Delta \tilde{I}_i$ have been calibrated using ensemble-averaged $\langle \tilde{I}_i(t)\rangle$ (see details in \cite{supplement}), giving in arbitrary units $\Delta \tilde{I}_z=4.0$ and $\Delta \tilde{I}_\varphi=4.4$. The offsets $\tilde{I}_i^{\rm off}$ are approximately zeroed individually for each trace by measuring the non-rotating physical qubit after each run. Additional offset removal, $|\tilde{I}_i^{\rm off}|\approx 0.15-0.20$, for all traces with the same $\varphi$ is done using $\langle \tilde{I}_i(t)\rangle$ \cite{supplement}. For calculating the correlators, we average over the ensemble of $\sim$200,000 traces and additionally average over time $t_1$ in Eq.\ (\ref{eq:corr-def}) within the 0.5 $\mu$s range $1\,\mu{\rm s}\leq t_1 \leq 1.5\, \mu{\rm s}$ (first 1 $\mu$s is not used to avoid transients in the experimental procedure, and longer averaging reduces the range for $\tau$; we also used averaging over 1 $\mu$s duration with similar results). Note that in the experiment the applied microwave phases in the two measurement channels actually correspond to angles $\pm \varphi_n/2$; however, because of rotational symmetry, we still label the first measured operator as $\sigma_z$ and the second operator as $\sigma_\varphi$. Also note that we use subscripts $z$ and $\varphi$ in various notations ($\tilde I_i$, $\kappa_i$, etc.) simply to distinguish the first (``$z$'') and second (``$\varphi$'') measurement channels.

Figure \ref{fig:theory-vs-exp}(a) shows the agreement between the theory and the experimental data, where the solid lines show the symmetrized cross-correlator $[K_{z\varphi}(\tau) +K_{\varphi z}(\tau)]/2$ calculated from the experimental traces for 11 values of the angle $\varphi$, while the dashed lines correspond to the theoretical result, Eq.\ (\ref{eq:K-zphi}).  For the analytics we used $\tilde\Omega_{\rm R}=0$; however, there is practically no dependence on $\tilde\Omega_{\rm R}$ for the symmetrized cross-correlator, since the dependence comes only via Eq.\ (\ref{eq:Gamma_pm}). Note that because of the Markovian assumption, our theory is formally valid only for $\tau\agt \kappa_i^{-1} \sim 30$ ns; however, the experimental results agree with the theory even at $\tau < \kappa_i^{-1}$. Figure \ref{fig:theory-vs-exp}(b) shows the same symmetrized cross-correlator at $\tau =0$ as a function of $\varphi$. The agreement between the theory ($\cos \varphi$, line) and the experiment (crosses) is also very good.

The self-correlator $K_{zz}(\tau)$ as a function of $\tau$ is shown in Fig.\ \ref{fig:theory-vs-exp}(c) for 11 values of $\varphi$ (results for $K_{\varphi\varphi}$ are similar).
The agreement between the theory (dashed lines) and experiment (solid lines) is good, except for small $\tau$ (discussed below). A significant discrepancy for values of $\varphi$ close to $\pi/2$ is probably caused by remaining offsets $\tilde{I}_i^{\rm off}$, which vary from trace to trace. Note that the lines in \ref{fig:theory-vs-exp}(c) come in pairs, corresponding to angles $\varphi_n$ and $\pi-\varphi_n$. The separation of the analytical lines in the pairs is due to $\delta \varphi$, while separation of experimental lines is smaller, probably indicating a smaller value of $\delta \varphi$ (partial compensation could be due to imperfect phase matching of applied microwaves or their dispersion in the cable).

Looking at the experimental self-correlators $K_{zz}(\tau)$ and $K_{\varphi\varphi}(\tau)$ at small $\tau$ for $\varphi_n=\pi/2$ [Fig.\ \ref{fig:theory-vs-exp}(d)], we see that in contrast to the theoretical results, there is a very significant increase of $K_{ii} (\tau)$ at $\tau \alt 0.1 \, \mu{\rm s}$. The discrepancy is due to the assumption of delta-correlated noise in our theory, while in the experiment the amplifying chain has a finite bandwidth (the Josephson parametric amplifiers have a half-bandwidth of 3.6 MHz and 10 MHz for $\sigma_z$ and $\sigma_\varphi$ channels, respectively), and the output signals $\tilde{I}_i(t)$ are also passed through analog filters with a quite sharp cutoff at $\sim$25 MHz (this cutoff produces clearly visible oscillations with $\sim$40 ns period). Therefore, the theoretical delta-function contribution $\tau_i\, \delta(\tau)$ to $K_{ii}(\tau)$ becomes widened in experiment. It is interesting to note that, somewhat counterintuitively, a finite bandwidth of measurement resonator modes does not produce a contribution to $K_{ii}(\tau)$ at $0<\tau \alt \kappa_i^{-1}$ when $\Gamma_i\ll \kappa_i$ \cite{supplement} ($\kappa_z^{-1}\approx 37$~ns, $\kappa_\varphi^{-1}\approx 22$~ns). This can be understood by considering a resonator without a qubit; then a finite bandwidth $\kappa_i$ does not affect the amplified delta-correlated vacuum noise, so that only classical fluctuations of the resonator field (e.g., due to parameter fluctuations or elevated resonator temperature) will produce output fluctuations with $2/\kappa_i$ time scale. We have checked that the lines in Fig.\ \ref{fig:theory-vs-exp}(d) do not contain noticeable exponential contributions with decay time of $2/\kappa_i$ (small expected contributions with amplitude on the order of $\Gamma_j/\kappa_i$ \cite{supplement} are below experimental accuracy).

\begin{figure}[tb]
\centering
\includegraphics[width=0.85\linewidth, trim = 2.0cm 2.0cm 2.0cm 1.9cm]{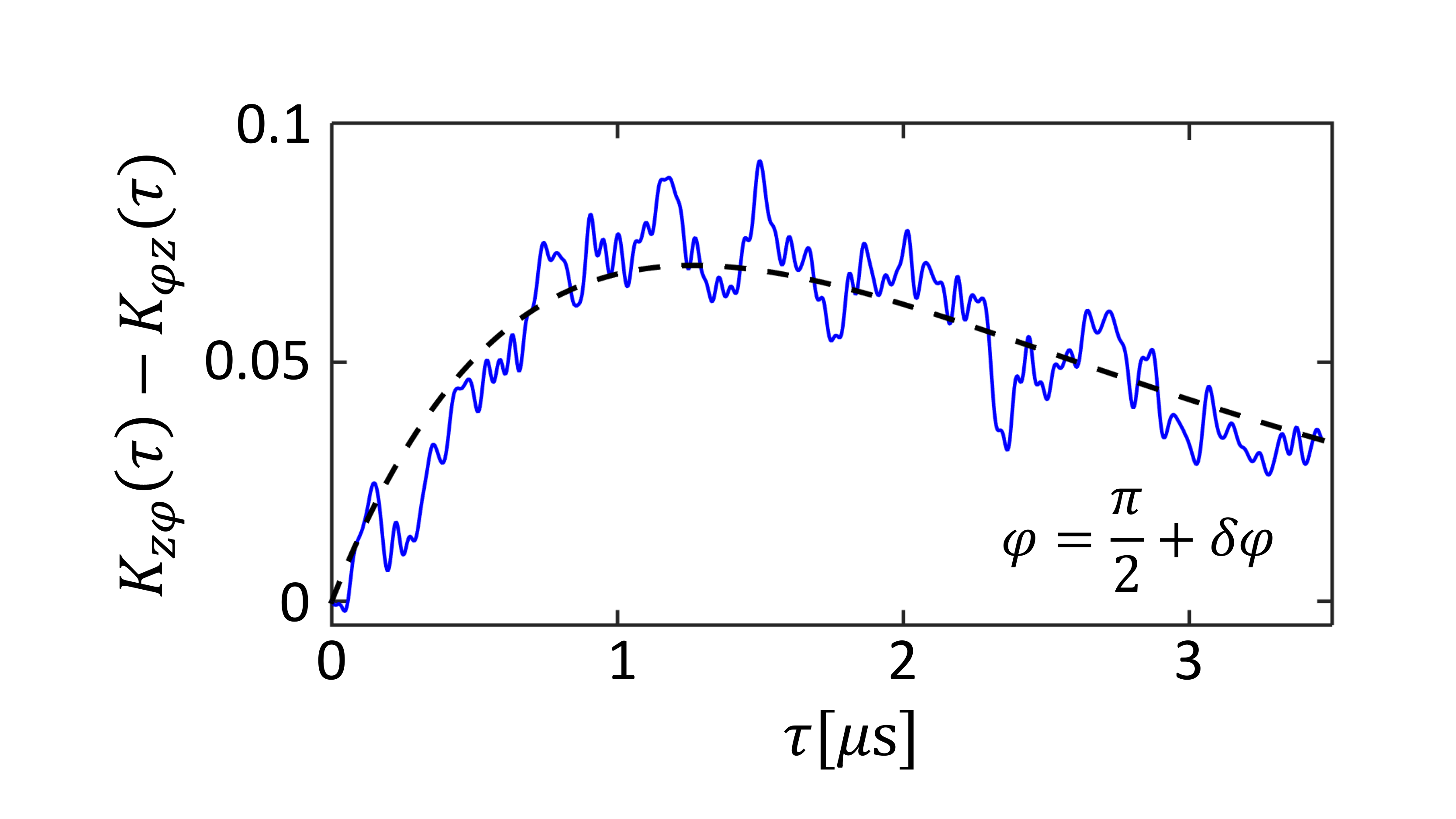}
\caption{Estimation of the residual Rabi frequency $\tilde\Omega_{\rm R}$ from the antisymmetrized cross-correlator $K_{z\varphi}(\tau)-K_{\varphi z}(\tau)$. Solid line shows experimental results for $\varphi_n=\pi/2$, while dashed line shows Eq.\ (\ref{eq:Kzphi-antisym}) with the fitted value $\tilde\Omega_{\rm R}/2\pi = 12$ kHz. Averaging over $\sim$200,000 experimental traces produces a clearly-visible difference signal, though with a significant noise.  }
 \label{fig:DeltaKzphi}
\end{figure}

{\it Estimation of residual $\tilde\Omega_{\rm R}$}.---We now show that the antisymmetrized cross-correlator is a useful tool and can be used to estimate small residual Rabi oscillations frequency $\tilde\Omega_{\rm R}$ in the experiment. From Eq.\ (\ref{eq:K-zphi}) we find
    \begin{align}
\label{eq:Kzphi-antisym}
 K_{z\varphi}(\tau)-K_{\varphi z}(\tau )=\frac{2\tilde\Omega_{\rm R}\sin\varphi}{\Gamma_+-\Gamma_-}\left(e^{-\Gamma_-\tau} - e^{-\Gamma_+\tau}\right) .
    \end{align}
Since in the case $|\tilde\Omega_{\rm R}|\ll \Gamma_{z,\varphi}$ we can neglect $\tilde\Omega_{\rm R}$ in Eq.\ (\ref{eq:Gamma_pm}) for $\Gamma_{\pm}$, Eq.\ (\ref{eq:Kzphi-antisym}) gives a direct way to find $\tilde\Omega_{\rm R}$ from the experimental antisymmetrized cross-correlator. The solid line in Fig.\ \ref{fig:DeltaKzphi} shows $K_{z\varphi}(\tau)-K_{\varphi z}(\tau )$ from the experimental data for $\varphi=\pi/2$. Fitting this dependence on $\tau$ with Eq.\ (\ref{eq:Kzphi-antisym}) (dashed line), we find the value $\tilde\Omega_{\rm R}/2\pi \approx 12$~kHz, which is within the experimentally expected range of frequency mismatch between $\Omega_{\rm R}$ and $\Omega_{\rm rf}$. Note that the overall shapes of the solid and dashed lines agree well with each other. Estimation of $\tilde\Omega_{\rm R}$ via the antisymmetrized cross-correlation is a very sensitive method and can be used to further reduce $|\tilde\Omega_{\rm R}|$ in an experiment, in which a direct measurement of 40 MHz Rabi oscillations with a few-kHz accuracy is a difficult task.

{\it Conclusion}.---Using the quantum Bayesian theory for a simultaneous measurement of non-commuting qubit observables, we obtained analytical results for the self- and cross-correlators of the output signals from the measurement. Their comparison with experimental results shows a very good agreement. The correlators can be used for sensitive parameter estimation, in particular, to estimate and eliminate the mismatch between the Rabi oscillations and the sideband frequency shift used for measurement.

{\it Acknowledgements}.---We thank Justin Dressel and Andrew Jordan for useful discussions. The work was supported by ARO grant No. W911NF-15-0496. L.S.M acknowledges support from the National Science Foundation Graduate Fellowship Grant No. 1106400.

\clearpage 
\onecolumngrid 
\vspace{\columnsep}
\begin{center}
\textbf{\large Supplemental material for ``Correlators in simultaneous measurement of non-commuting qubit observables''}
\end{center}
\vspace{\columnsep}
\twocolumngrid

\setcounter{equation}{0}
\setcounter{figure}{0}
\setcounter{table}{0}
\setcounter{page}{1}

\renewcommand{\theequation}{S\arabic{equation}}
\renewcommand{\thefigure}{S\arabic{figure}}
\renewcommand{\bibnumfmt}[1]{[S#1]}
\renewcommand{\citenumfont}[1]{S#1}

\title{Supplemental material for ``Correlators in simultaneous measurement of non-commuting qubit observables''}
%

%

\section{Experimental setup}

The experimental setup is the same as the one used in the experiment \cite{S-Shay2016}, where full details can be found. For clarity we briefly describe the experimental apparatus for simultaneously applying and controlling two measurement observables. We use a transmon qubit placed inside an aluminum cavity, such that it is dispersively coupled to the two lowest modes of the cavity. The cavity has two outputs, each primarily coupled to a different mode. The outputs of these modes are amplified using two lumped-element Josephson parametric amplifiers (LJPA) operated in phase sensitive mode. Each mode is then used to measure an observable of the qubit, as described below. The apparatus is cooled to 30 mK inside a dilution refrigerator.

We drive Rabi oscillations $\Omega_{\rm R}/2\pi = 40 ~\mathrm{MHz}$ on the qubit by applying a resonant microwave tone modulated by an arbitrary waveform generator.
In the frame rotating with $\Omega_{\rm R}$, this produces an effective low frequency qubit. To couple the effective qubit to the cavity modes for measurement, we apply a pair of microwave sidebands to each mode. The sidebands are detuned above and below the two cavity modes by $\Omega_R$, which leads to a resonant interaction between the qubit Rabi oscillations and the mode. This coupling may be understood as a stroboscopic measurement of the qubit oscillations. The relative phase of the sidebands determines which quadrature of the qubit oscillations is measured. This coupling causes the cavity mode state to displace in a way that depends on the state of the qubit.  We couple to the internal cavity field using a small antenna that protrudes into the cavity, allowing read out the cavity state as described above. Quantum trajectory reconstructions are validated using post-selection and tomographic measurements.

\section{Quantum Bayesian approach to qubit measurement in Rabi-rotated frame }

In this section we develop the quantum Bayesian theory of the stroboscopic qubit measurement in the Rabi-rotated frame, used in the experiment \cite{S-Shay2016} and briefly described above. We start with measurement of one effective observable $\sigma_\varphi=\sigma_z \cos\varphi +\sigma_x \sin \varphi$, then adding the second measurement in the same way and deriving Eqs.\ (4)--(6) of the main text. In this derivation we assume that the qubit Rabi frequency $\Omega_{\rm R}$ is exactly equal to the sideband frequency shift $\Omega_{\rm rf}$ (which defines the rotating frame), while a small mismatch between $\Omega_{\rm R}$ and $\Omega_{\rm rf}$ is added later via Eq.\ (7) of the main text (also discussed in this section). The focus is on the simple physics of the qubit measurement in the rotating frame.

\subsection{Measurement of one observable $\sigma_\varphi$}

The {\it physical} qubit is Rabi-rotated with frequency $\Omega_{\rm R}$ about the $y$-axis, so that its Bloch coordinates rotate as
    \begin{eqnarray}
&&      x_{\rm ph}(t)= r_0 \sin (\Omega_{\rm R} t +\phi_0),
    \label{x-phys}\\
&&      y_{\rm ph}(t)=y_0,
    \label{y-phys} \\
&& z_{\rm ph}(t) =r_0 \cos (\Omega_{\rm R} t+\phi_0),
        \label{z-phys}
    \end{eqnarray}
where the radius $r_0(t)$ within the $xz$-plane, the rotation phase $\phi_0(t)$, and the coordinate $y_0(t)$ slowly change in time (e.g., due to measurement). The oscillations of the qubit $z$-component lead to a small change of the effective resonator frequency,
    \be
    \omega_{\rm r}(t) = \omega_{\rm r}^{\rm m} +\chi \, r_0 \cos (\Omega_{\rm R} t+\phi_0),
    \ee
where  $\chi$  is the (small) dispersive coupling between the qubit and the measurement resonator mode, and $\omega_{\rm r}^{\rm m}$ is the mean value between the resonator frequencies for the physical qubit states $|0\rangle$ and $|1\rangle$. Note that in this derivation, fast-oscillating $\omega_{\rm r}(t)$ is the value averaged over the physical qubit states, and we neglect quantum backaction developing during short time scale $\Omega_{\rm R}^{-1}$.

The sideband drive of the resonator at frequencies $\,\,\, \omega_{\rm d} \pm \Omega_{\rm rf}$ with equal amplitudes (here for simplicity we assume $\omega_{\rm d}=\omega_{\rm r}^{\rm m}$ and $\Omega_{\rm rf}=\Omega_{\rm R}$), produces the Hamiltonian term
    \be
    H_{\rm drive}/\hbar =   \varepsilon  \sin (\Omega_{\rm R} t+\varphi)\, a^\dagger +{\rm h.c.},
    \label{H-drive}\ee
where $\varepsilon$ is the normalized amplitude, $\varphi$ depends on the initial phase shift between the sideband tones, $a^\dagger$ is the creation operator for the resonator, and we use the rotating frame based on $\omega_{\rm d}$. The form of this term follows from the usual trigonometric relation for adding the sideband tones, $(\varepsilon/2)\sin[(\omega_{\rm d} + \Omega_{\rm R})t+\varphi]-(\varepsilon/2)\sin[(\omega_{\rm d} - \Omega_{\rm R})t-\varphi]=\varepsilon \sin(\Omega_{\rm R}t+\varphi)\cos(\omega_{\rm d} t)$.

This leads to the following dynamics of the resonator's coherent state  $|\alpha (t)\rangle$ [or classical field $\alpha (t)$] in the rotating frame based on $\omega_{\rm d}$,
    \be
\dot{\alpha} = -i \chi r_0 \cos(\Omega_{\rm R} t+\phi_0) \, \alpha - i \varepsilon \sin (\Omega_{\rm R} t+\varphi ) -\frac{\kappa}{2}\alpha,
    \label{alpha-dot}\ee
where we also took into account the resonator damping with energy decay rate $\kappa$.

Now let us solve the evolution equation (\ref{alpha-dot}), assuming $\kappa \ll \Omega_{\rm R}$ and $|\chi| \ll \Omega_{\rm R}$. The drive term produces fast oscillation of $\alpha$ with Rabi frequency $\Omega_{\rm R}$,
    \be
\Delta \alpha (t)= i \frac{\varepsilon}{\Omega_{\rm R}} \cos (\Omega_{\rm R} t +\varphi ).
    \label{Delta-alpha} \ee
Inserting this oscillation into the first term of Eq.\ (\ref{alpha-dot}), using the trigonometric formula $\cos(\Omega_{\rm R} t+\phi_0)\cos (\Omega_{\rm R} t+\varphi)= \frac{1}{2}\cos(\phi_0-\varphi)+\frac{1}{2}\cos (2\Omega_{\rm R} t + \phi_0 +\varphi )$, and neglecting oscillations with frequency $2\Omega_{\rm R}$, we obtain the equation for the slow evolution,
    \begin{eqnarray}
&&\dot{\alpha}_{\rm s} =   \frac{\chi\varepsilon}{2\Omega_{\rm R}} \, r_0 \cos(\phi_0 -\varphi )  -\frac{\kappa}{2}\alpha_{\rm s} ,
    \label{alpha-dot-2}\\
&& \alpha (t) =\alpha_s (t)+\Delta \alpha(t).
    \end{eqnarray}
Note that we can neglect the additional fast oscillations produced by the first term in Eq.~(\ref{alpha-dot}), $\Delta_2\alpha (t)=-i(\chi/\Omega_{R}) r_0\sin(\Omega_{\rm R} t+\phi_0)\,\alpha_{\rm s}$, in comparison with (\ref{Delta-alpha}), because we assume $\chi^2/(\Omega_{\rm R}\kappa)\ll 1$.

We see that the evolution (\ref{alpha-dot-2}) of the resonator field $\alpha_{\rm s}$ depends on the state of the {\it effective} qubit,
    \be
  x =r_0 \sin(\phi_0), \,\,\, y=y_0, \,\,\,   z=r_0 \cos (\phi_0),
    \ee
which corresponds to the physical qubit (\ref{x-phys})--(\ref{z-phys}) in the rotating frame $\Omega_{\rm R}$. Moreover, this dependence is only on the Bloch $\varphi$-coordinate of the effective qubit, which is within the $xz$-plane at an angle $\varphi$ from the $z$-axis, we see this since
    \be
    r_0 \cos(\phi_0 -\varphi ) = {\rm Tr}[\sigma_{\varphi} \rho (t)],
    \ee
where $\rho(t)$ is the slowly-varying density matrix of the effective qubit.

At this stage, we make use of the quantum Bayesian approach \cite{S-Korotkov-99-01,S-Korotkov-2011,S-Korotkov-2016} to describe the qubit evolution due to measurement. Since the oscillating part $\Delta \alpha$ of the resonator field [Eq.\ (\ref{Delta-alpha})] does not depend on the qubit state, it can be neglected in the analysis. In contrast, homodyne measurement of the leaked field $\alpha_{\rm s}$
gives us information on the value of the $\varphi$-coordinate of the effective qubit, which is a two-level system similar to the physical qubit. Inevitably, this information gradually collapses the effective qubit, i.e., changes its state according to the acquired information.

The two $\sigma_\varphi$-basis states $|1_\varphi\rangle$ and $|0_\varphi\rangle$  of the effective qubit ($\sigma_\varphi |1_\varphi\rangle =|1_\varphi\rangle$, $\sigma_\varphi |0_\varphi\rangle =-|0_\varphi\rangle$) produce two steady states of the resonator, respectively (excluding oscillating $\Delta\alpha$),
    \be
    \alpha_{\rm st,1} =  \frac{\chi \varepsilon}{\Omega_{\rm R}\kappa }, \,\,\, \alpha_{\rm st, 0}=-\alpha_{\rm st,1}.
    \label{alpha-st}\ee
This is all what is needed in the Markovian ``bad cavity'' regime (when the evolution of the effective qubit is much slower than $\kappa$), which is assumed in the main text. Since in circuit QED only the difference between $\alpha_{\rm st,1}$ and $\alpha_{\rm st,0}$ is important for the analysis of the qubit evolution due to measurement in the ``bad cavity'' regime  \cite{S-Wiseman1993, S-Gambetta2008, S-Korotkov-2011, S-Korotkov-2016}, the situation is {\it equivalent} to the qubit evolution due to measurement in the standard setup \cite{S-Blais-2004} with the same  $\alpha_{\rm st,1}-\alpha_{\rm st,0}$. Correspondingly, the quantum Bayesian formalism in the ``bad cavity'' regime is exactly the same as for the standard circuit QED setup \cite{S-Korotkov-2011}, which coincides with the Bayesian formalism for qubit measurement using a quantum point contact \cite{S-Korotkov-99-01}. The only difference is that now we discuss the evolution of the effective qubit instead of the physical qubit.

In particular, when a phase-sensitive amplifier is used, the response $\Delta\tilde I$ of the detector to the effective qubit state has the dependence $\Delta\tilde I=\Delta\tilde{I}_{\rm max}\cos \theta$ on the phase difference $\theta$ between the amplified quadrature (at the microwave frequency $\omega_{\rm d}$) and the optimal quadrature [which is real (horizontal), as follows from Eq.\ (\ref{alpha-st})]. The informational backaction is proportional to $\cos \theta$, while phase backaction is proportional to $\sin \theta$, with ensemble dephasing of the effective qubit,
    \be
\Gamma=\frac{\kappa}{2} \, |\alpha_{\rm st,1}-\alpha_{\rm st,0}|^2,
    \ee
not depending on $\theta$ \cite{S-Gambetta2008,S-Korotkov-2011, S-Korotkov-2016}. Evolution of the effective qubit state due to measurement is given by Eqs.\ (18) and (25) of Ref.\ \cite{S-Korotkov-2011} in the basis $\{|0_\varphi\rangle, |1_\varphi\rangle\}$  (Eqs.\ (12) and (13) of Ref.\ \cite{S-Korotkov-2016}). In this basis, the diagonal matrix elements $\rho_{00}$ and $\rho_{11}$ evolve due to the classical Bayes rule, while the off-diagonal elements $\rho_{01}$ and $\rho_{10}$ evolve due to evolving product $\rho_{00}\rho_{11}$ and also due to phase backaction.

In the above derivation we assumed an exactly resonant microwave drive, $\omega_{\rm d}=\omega_{\rm r}^{\rm m}$. If this is not the case, $|\omega_{\rm d}-\omega_{\rm r}^{\rm m}| \sim\kappa$, then there will be an extra term $-i (\omega_{\rm r}^{\rm m}-\omega_{\rm d})\alpha$ in Eq.\ (\ref{alpha-dot}), which will lead to an extra term $-i (\omega_{\rm r}^{\rm m}-\omega_{\rm d})\alpha_{\rm s}$ in Eq.\ (\ref{alpha-dot-2}). Correspondingly, the steady states of the field $\alpha_{\rm s}$ for the effective qubit states $|1_\varphi\rangle$ and $|0_\varphi\rangle$  are
    \be
    \alpha_{\rm st,1} =-\alpha_{\rm st, 0} =  \frac{\chi \varepsilon}{\Omega_{\rm R}[\kappa +2i(\omega_{\rm r}^{\rm m}-\omega_{\rm d})] }
    \label{alpha-st-2}\ee
instead of Eq.\ (\ref{alpha-st}), so that the optimal quadrature is no longer horizontal (real). The quantum Bayesian formalism remains the same.

If the effective rotating-frame qubit is measured by only one detector ($\sigma_\varphi$, but no $\sigma_z$) and $\Omega_{\rm R}=\Omega_{\rm rf}$, then it is possible to go beyond the ``bad cavity'' limit and analyze transients within the time scale $\kappa^{-1}$. The derivation of the quantum Bayesian formalism for this case exactly follows the derivation in Ref.\ \cite{S-Korotkov-2016} and uses the field evolution equation (\ref{alpha-dot-2}) instead of the steady-state solution~(\ref{alpha-st}).

\subsection{Derivation of Eqs.\ (4)--(6) in the main text }

In the absence of phase back-action, the quantum Bayesian equations describing continuous measurement of qubit $\sigma_z$ observable in the Markovian approximation are \cite{S-Korotkov-99-01,S-Korotkov-2016}
\begin{align}
\dot x=&\,- \tau_z^{-1}  xz \, I_z(t)    - \gamma_z x,
    \label{S-Strat-x} \\
\dot y=&\, - \tau_z^{-1} yz \, I_z(t) - \gamma_z y ,
    \label{S-Strat-y} \\
\dot z= &\,  \tau_z^{-1}(1-z^2)\, I_z(t) ,
\label{S-Strat-z}
\end{align}
in the Stratonovich form, where
    \be
    I_z(t)= {\rm Tr}[\sigma_z \rho (t)]+\sqrt{\tau_z} \, \xi_z (t)
    \label{S-I-z}\ee
is the normalized output signal, $\xi_z(t)$ is the normalized white noise, $\langle \xi_z(t) \, \xi_z(t')\rangle =\delta (t-t')$, $\tau_z$ is the ``measurement'' time after which the signal-to-noise ratio reaches 1, the qubit density matrix is $\rho = (\openone +x\sigma_x+ y\sigma_y +z\sigma_z)/2$, and qubit dephasing $\gamma_z$ in individual measurement is related to the ensemble dephasing $\Gamma_z$ as $\gamma_z=\Gamma_z-(2\tau_z)^{-1}$.

In the It\^o form (i.e., using the forward definition of derivatives instead of the symmetric definition) these evolution equations become \cite{S-Korotkov-99-01,S-Korotkov-2016}
\begin{align}
\dot x=&\, - \tau_z^{-1/2}xz \, \xi_z  - \Gamma_z x  ,
    \label{S-Ito-x} \\
\dot y=&\, - \tau_z^{-1/2}yz \, \xi_z  - \Gamma_z y  ,
    \label{S-Ito-y} \\
\dot z=&\, \tau_z^{-1/2}(1-z^2)\, \xi_z .
    \label{S-Ito-z}
\end{align}

When the observable $\sigma_\varphi$ is measured instead of $\sigma_z$, these equations remain the same \cite{S-Ruskov2010} in the basis of eigenstates $|0_\varphi\rangle$ and $|1_\varphi\rangle$, so that we can simply  change the notation: $x\to x_\varphi$, $y\to y_\varphi$, $z\to z_\varphi$. Rotating back to the usual basis, i.e., using the transformation $x=x_\varphi \cos \varphi + z_\varphi \sin \varphi$ , $y=y_\varphi$, $z=z_\varphi \cos\varphi -x_\varphi \sin \varphi$, we obtain for the It\^o form
\begin{align}
\dot x=&\, - \tau_\varphi^{-1/2}\left[xz\cos\varphi - (1-x^2)\sin\varphi\right] \xi_\varphi ,
    \nonumber \\
&\, - \Gamma_{\varphi}  \cos\varphi\left(x\cos\varphi - z\sin\varphi\right) ,
    \label{S-Ito-x-phi} \\
\dot y=&\, - \tau_\varphi^{-1/2} y \left[z\cos\varphi + x\sin\varphi\right] \xi_\varphi  - \Gamma_\varphi \, y ,
    \label{S-Ito-y-phi} \\
\dot z=&\, \tau^{-1/2}_\varphi\left[(1 - z^2)\cos\varphi -xz\sin\varphi\right]\xi_\varphi
    \nonumber \\
          &\, + \Gamma_{\varphi}  \sin\varphi \left(x\cos\varphi - z\sin\varphi\right) ,
    \label{S-Ito-z-phi}
    \end{align}
with
    \be
    I_\varphi(t)= {\rm Tr}[\sigma_\varphi \rho (t)]+\sqrt{\tau_\varphi} \, \xi_\varphi (t).
    \label{S-I-varphi}\ee

When both $\sigma_z$ and $\sigma_\varphi$ measurements are performed at the same time, we simply add the terms from Eqs.\ (\ref{S-Ito-x})--(\ref{S-Ito-z}) and (\ref{S-Ito-x-phi})--(\ref{S-Ito-z-phi}) \cite{S-Ruskov2010} (with uncorrelated noises $\xi_z$ and $\xi_\varphi$ in the two channels), thus obtaining Eqs.\ (4)--(6) of the main text.

\subsection{Correction to measured rotation phase $\varphi$}

As was discussed above, the phase $\varphi$ in the double-sideband drive $\varepsilon \sin (\Omega_{\rm R} t+\varphi)\cos (\omega_{\rm d}t)$ [see Eq.\ (\ref{H-drive})] directly determines the angle for the measured operator $\sigma_\varphi$ for the effective qubit. This followed from the approximate solution of Eq.\ (\ref{alpha-dot}). As we will see below, a more accurate solution shows a small correction to the measured direction $\varphi$.

Neglecting the first term in Eq.\ (\ref{alpha-dot}) but still keeping the last term, we obtain the (exact) oscillating solution
   \be
    \Delta \alpha (t) = \frac{i\varepsilon}{\Omega_{\rm R}+\frac{\kappa^2}{4\Omega_{\rm R}}}
    \left[ \cos(\Omega_{\rm R}t+\varphi) -\frac{\kappa}{2\Omega_{\rm R}} \sin (\Omega_{\rm R}t+\varphi) \right] ,
    \label{Delta-alpha-better}\ee
which is more accurate than Eq.\ (\ref{Delta-alpha}). (Note that the additional term $\propto e^{-\kappa t/2}$ is naturally included in the slow dynamics.)
Inserting $\Delta \alpha (t)$ into the first term of Eq.\ (\ref{alpha-dot}), we obtain the following evolution of the slow part $\alpha_{\rm s}$ of the total field $\alpha =\alpha_{\rm s}+\Delta \alpha$:
    \begin{align}
\dot{\alpha}_{\rm s} = & \,  \frac{\chi\varepsilon}{2\left( \Omega_{\rm R}+\frac{\kappa^2}{4\Omega_{\rm R}}\right)} \, r_0 \left[ \cos(\phi_0 -\varphi )-\frac{\kappa}{2\Omega_{\rm R}}\sin (\varphi -\phi_0) \right]
    \nonumber \\
& -\kappa \alpha_{\rm s} /2 ,
    \label{alpha-dot-3}\end{align}
which in the case $\kappa/\Omega_{\rm R} \ll 1$ is approximately
        \be
\dot{\alpha}_{\rm s} =   \frac{\chi\varepsilon}{2\Omega_{\rm R}} \, r_0 \cos(\phi_0 -\varphi -\kappa/2\Omega_{\rm R} )  -\frac{\kappa}{2}\alpha_{\rm s} .
    \label{alpha-dot-4}\ee
This equation coincides with Eq.\ (\ref{alpha-dot-2}), except $\varphi$ is replaced with $\varphi+\kappa/2\Omega_{\rm R}$, thus slightly changing the direction of  the measured operator for the effective qubit,
   \be
    r_0 \cos(\phi_0 -\varphi -\kappa/2\Omega_{\rm R} ) = {\rm Tr}[\sigma_{\varphi+\kappa/2\Omega_{\rm R}} \rho (t)].
    \ee

In the case when two channels simultaneously measure nominal operators $\sigma_z$ and $\sigma_\varphi$ (with corresponding resonator bandwidths $\kappa_z$ and $\kappa_\varphi$), the measured directions on the Bloch sphere are actually $\kappa_z/2\Omega_{\rm R}$ and $\varphi+\kappa_\varphi/2\Omega_{\rm R}$, so that the relative angle is $\varphi+\delta \varphi$ with
    \be
    \delta \varphi = \frac{\kappa_\varphi-\kappa_z}{2\Omega_{\rm R}} .
    \ee
This correction of $\sim2^\mathrm{o}$ was used in the main text when we compared the experimental results with the theory results.

\subsection{Decoherence of the effective qubit}

Now let us derive Eq.\ (7) of the main text, describing the evolution of the effective qubit not related to measurement.

The evolution of the {\it physical} qubit due to Rabi oscillations and environmental decoherence (energy relaxation and pure dephasing) is described by the standard master equation
\begin{eqnarray}
\label{phys-qubit-EOM}
&& \dot \rho_{\rm ph} = \frac{-i}{\hbar}\,[H_{\rm ph},\rho_{\rm ph}] + \frac{1}{2T_{\rm pd}}\,\mathcal{L}[{\sigma_z}] + \frac{1}{T_1}\,\mathcal{L}[{\sigma_-}], \qquad
    \\
&& \mathcal{L}[A] \equiv  A\rho_{\rm ph}A^\dagger - \frac{1}{2}\left(A^{\dagger}A\rho_{\rm ph} - \rho_{\rm ph} A^{\dagger}A\right), \label{Lindblad}
\end{eqnarray}
where $\rho_{\rm ph}(t)$ and $H_{\rm ph} = \hbar\Omega_{\rm R}\sigma_y/2$ are the density matrix and Hamiltonian of the physical qubit, respectively, the Lindblad-operator evolution with $A=\sigma_z$ describes pure dephasing, and energy relaxation corresponds to  $A=\sigma_- =|0\rangle \langle 1|= (\sigma_x - i\sigma_y)/2$.

To convert Eq.~\eqref{phys-qubit-EOM} into the rotating frame (with frequency $\Omega_{\rm rf}$), we apply the unitary transformation $\rho(t) = U^\dagger(t)\, \rho_{\rm ph}(t)\,U(t)$, where $\rho(t)$ is the effective qubit density matrix and $U(t) =\exp(-i\Omega_{\rm rf}t\sigma_y/2)$. This gives
\begin{align}
\dot x &= \tilde\Omega_{\rm R}z - \left(\frac{1}{2T_{\rm pd}} + \frac{3}{4T_1} \right)x + \frac{1}{T_1}\sin(\Omega_{\rm rf}t)
    \nonumber \\
&\; + \left( \frac{1}{4T_1} - \frac{1}{2T_{\rm pd}} \right)\left[x\cos(2\Omega_{\rm rf}t) + z\sin(2\Omega_{\rm rf}t)\right],
    \label{S-env-decoh-x}  \\
\dot y &= -\left(\frac{1}{T_{\rm pd}} + \frac{1}{2T_1}\right)y,
    \label{S-env-decoh-y} \\
\dot z &= -\tilde \Omega_{\rm R}x - \left( \frac{1}{2T_{\rm pd}} + \frac{3}{4T_1} \right) z - \frac{1}{T_1}\cos(\Omega_{\rm rf}t)
    \nonumber \\
& \; + \left( \frac{1}{4T_1} - \frac{1}{2T_{\rm pd}} \right)\left[x\sin(2\Omega_{\rm rf}t) - z\cos(2\Omega_{\rm rf}t)\right],
    \label{S-env-decoh-z} \end{align}
where $\tilde \Omega_{\rm R} = \Omega_{\rm R} - \Omega_{\rm rf}$.
Since $\Omega_{\rm rf}$ is much faster than the evolution of the effective qubit,  in Eqs.\ (\ref{S-env-decoh-x}) and (\ref{S-env-decoh-z}) we can neglect the oscillating terms. Finally expressing $T_{\rm pd}$ via $T_2$ and $T_1$, so that $T_{\rm pd}^{-1}+ (2T_1)^{-1}=T_2^{-1}$ and $(2T_{\rm pd})^{-1}+3(4T_1)^{-1}=
(T_{1}^{-1}+T_2^{-1})/2$, we obtain
\begin{align}
\dot x &= \tilde\Omega_{\rm R}z - \frac{1}{2}\left(T_1^{-1} + T_2^{-1}\right) x,
    \label{S-decoh-x} \\
\dot y &= T_2^{-1}y,
     \label{S-decoh-y} \\
\dot z &= -\tilde\Omega_{\rm R}x - \frac{1}{2}\left(T_1^{-1} + T_2^{-1}\right)z,
     \label{S-decoh-z}\end{align}
which is Eq.\ (7) of the main text.

Since Eqs.\ (\ref{S-env-decoh-x})--(\ref{S-env-decoh-z}) describe only the evolution not related to measurement, while Eqs.\ (4)--(6) of the main text  describe only the evolution due to measurement, we need to simply add terms in these equations to describe the combined evolution of the effective qubit.
We would like to emphasize that the derivation presented here relies on a significant separation of frequency scales
\be
\big(T_1^{-1},T_2^{-1},|\tilde \Omega_{\rm R}|\big)\ll \big(\Gamma_z, \Gamma_\varphi\big) \ll \big(\kappa_{z},\kappa_{\varphi}\big)\ll \Omega_{\rm R} \label{S-ineq}.
\ee
In our experiment these inequalities are well satisfied:
$T_1=60\,\mu$s, $T_2=30\,\mu$s, $|\tilde\Omega_{\rm R}^{-1}|\agt 10 \, \mu$s,
$\Gamma_z^{-1}=\Gamma_\varphi^{-1}=1.31\,\mu {\rm s}$, $\kappa_z^{-1}= 37$~ns,  $\kappa_\varphi^{-1}=22.1$~ns, and $\Omega_{\rm R}^{-1}=4$~ns. The frequencies of the resonator modes should obviously be much larger than $\Omega_{\rm R}$; in our experiment $\omega_{{\rm r}, z}/2\pi=7.4\, {\rm GHz}$ and $\omega_{{\rm r}, \varphi}/2\pi=6.7\, {\rm GHz}$.

\section{Analytical results for correlators}

We will first derive Eqs.\ (14)--(16) of the main text for the correlators using the ``collapse recipe'' and then we discuss the derivation of this recipe from the quantum Bayesian equations and its correspondence to the quantum regression approach.

\subsection{Derivation of Eqs.\ (14)--(16) in the main text using ``collapse recipe''}

The collapse recipe for calculation of the correlators for the output signals (in the absence of phase backaction) was introduced in Ref.\ \cite{S-Korotkov2001sp}. It says that in order to calculate the ensemble-averaged correlator,
\begin{align}
\label{S-corr-def}
K_{ij}(\tau) \equiv  \langle I_j(t_1+\tau)\, I_i(t_1)\rangle,\,\,\,\,\, \tau>0,
\end{align}
we can replace the continuous measurement of $\sigma_i$ at the earlier time moment $t_1$ with its projective measurement. It is also possible to replace the continuous measurement of $\sigma_{j}$ at the later time moment $t_1+\tau$ with its projective measurement, but this is rather obvious and not important, since average values for the continuous and projective measurements coincide. In the following section we will show how this recipe can be derived from the quantum Bayesian equations (essentially repeating the derivation in Ref.\ \cite{S-Korotkov2001sp}); here we just use this recipe.

If the qubit state at time $t_1$ (it would be more accurate to say, right before $t_1$) is $\rho (t_1)$, then the projective measurement of $\sigma_i$ would produce the measurement result $I_i(t_1)=1$ with probability $\{1+{\rm Tr}[\sigma_i \rho (t_1)] \}/2$ and the result $I_i(t_1)=-1$ with probability $\{1-{\rm Tr}[\sigma_i \rho (t_1)] \}/2$. After this projective measurement, the qubit state is collapsed to the eigenstate $|1_i\rangle$ or $|0_i\rangle$, correspondingly. Ensemble-averaged evolution after that is simple, since for an ensemble a continuous measurement is equivalent to decoherence. If the state was collapsed to $|1_i\rangle$, then the further ensemble-averaged qubit evolution $\rho_{\rm av}(t |1_i)$ starts with $\rho_{\rm av}(t_1|1_i)=|1_i\rangle \langle 1_i|$. Then the average result of $\sigma_j$ measurement at time $t=t_1+\tau$ will be ${\rm Tr}[\sigma_j \rho_{\rm av} (t_1+\tau|1_i)]$, which will produce contribution ${\rm Tr}[\sigma_j \rho_{\rm av} (t_1+\tau|1_i)] \times 1$ to the correlator (\ref{S-corr-def}) with probability $\{1+{\rm Tr}[\sigma_i \rho (t_1)] \}/2$.  Similarly, the contribution corresponding to the state collapse to $|0_i\rangle$ at time $t_1$, is ${\rm Tr}[\sigma_j \rho_{\rm av} (t_1+\tau|0_i)] \times (-1)$ with probability $\{1-{\rm Tr}[\sigma_i \rho (t_1)] \}/2$. Summing these two cases, we obtain
    \begin{align}
& K_{ij}(\tau) = {\rm Tr}[\sigma_j \, \rho_{\rm av}(t_1+\tau|1_i)]\, \frac{1+ {\rm Tr}[\sigma_i\, \rho(t_1)]}{2}
    \nonumber \\
 &\hspace{1.3cm}-  {\rm Tr}[\sigma_j \, \rho_{\rm av}(t_1+\tau|0_i)] \, \frac{1- {\rm Tr}[\sigma_i \,\rho(t_1)]}{2} ,
    \label{S-Kij}\end{align}
which is Eq.\ (10) of the main text.

Next, we need to find $\rho_{\rm av}(t_1+\tau|1_i)$ and $\rho_{\rm av}(t_1+\tau|0_i)$. The ensemble-averaged evolution of the qubit is described by Eqs.\ (11)--(13) of the main text. It is easy to see that the evolution of the $y$-coordinate is decoupled and has the simple solution,
    \be
  y_{\rm av}(t_1+\tau) = e^{-(\Gamma_z+\Gamma_\varphi)\tau}\, y(t_1),
    \label{ens-y-sol}\ee
where we use the subscript ``av'' to indicate ensemble averaging. The evolution equations for the coordinates $x_{\rm av}$ and $z_{\rm av}$ can be written in the matrix form,
    \be
     \frac{d}{dt} \binom{x_{\rm av}}{z_{\rm av}} = {\bf M} \binom{x_{\rm av}}{z_{\rm av}},
        \label{ens-evol-matrix} \ee
     \be
     {\bf M} =  \begin{bmatrix} -(\Gamma_z +\cos^2\varphi \Gamma_\varphi+\gamma)\,\,\, &
   \sin\varphi \cos\varphi \, \Gamma_\varphi +\tilde\Omega_{\rm R} \\ \sin\varphi \cos\varphi \, \Gamma_\varphi -\tilde\Omega_{\rm R} & -(\sin^2\varphi\,\Gamma_\varphi +\gamma) \end{bmatrix}.
    \label{M-matrix}\ee
Diagonalizing the matrix $\bf M$, we find the solution
    \begin{eqnarray}
     && \hspace{-0.0cm} \binom{x_{\rm av}(t_1+\tau)}{z_{\rm
av}(t_1+\tau)} = \bigg[\frac{e^{-\Gamma_-\tau} +e^{-\Gamma_+\tau}}{2} \,  \openone +
 \frac{e^{-\Gamma_-\tau} - e^{-\Gamma_+\tau}}{2}
    \nonumber \\
&& \times \bigg( \frac{\Gamma_\varphi\sin 2\varphi +2\tilde{\Omega}_{\rm R} }{\Gamma_+-\Gamma_-}\, \sigma_x
  - \frac{\Gamma_z + \Gamma_\varphi\cos 2\varphi}{\Gamma_+-\Gamma_-} \, \sigma_z
 \bigg)
  \bigg]\binom{x(t_1)}{z(t_1)},
    \nonumber \\
    \label{S-evol-xz}\end{eqnarray}
where $\Gamma_{\pm}$  are the  eigenvalues of the matrix $-{\bf M}$, given by Eq.\ (16) of the main text and repeated here,
    \begin{align}
& \hspace{0cm} \Gamma_{\pm} =\frac{\Gamma_z +\Gamma_\varphi \pm \big[ \Gamma_z^2 + \Gamma_\varphi^2 +2\Gamma_z\Gamma_\varphi\cos(2\varphi) -4 \tilde\Omega_{\rm R}^2\big]^{1/2} }{2}
    \nonumber \\
&\hspace{0.9cm}  +  \gamma .
    \label{S-Gamma_pm} \end{align}

Note that our evolution is symmetric under the inversion operation $x\to -x, \, y\to -y, \, z\to -z$ (it is a unital map), and therefore
    \be
    {\rm Tr}[\sigma_j \, \rho_{\rm av}(t_1+\tau|0_i)] = -{\rm Tr}[\sigma_j \, \rho_{\rm av}(t_1+\tau|1_i)] .
    \ee
In this case Eq.\ (\ref{S-Kij}) simplifies to
    \be
 K_{ij}(\tau) = {\rm Tr}[\sigma_j \, \rho_{\rm av}(t_1+\tau|1_i)] ,
    \label{S-Kij-2}\ee
which no longer depends on the initial state $\rho (t_1)$.

In particular, if $\sigma_i=\sigma_z$, then to find $K_{ij}(\tau)$  we need to use initial conditions $x(t_1)=0$ and $z(t_1)=1$ in Eq.\ (\ref{S-evol-xz}), which gives
   \begin{eqnarray}
&& x_{\rm av}(t_1+\tau |1) =
\,\, \frac{\sin(2\varphi)\, \Gamma_\varphi+2\tilde{\Omega}_{\rm R}}{2(\Gamma_+-\Gamma_-)}   \left( e^{-\Gamma_-\tau} - e^{-\Gamma_+\tau}\right) , \;\;\; \quad
   \label{ens-x-sol} \\
&& z_{\rm av}(t_1+\tau |1 ) = \frac{1}{2} \left[ 1+\frac{\Gamma_z+\cos(2\varphi)\,\Gamma_\varphi}{\Gamma_+ -
\Gamma_-}\right]    e^{-\Gamma_-\tau }
    \nonumber \\
&& \hspace{2.1cm}   +\, \frac{1}{2} \left[ 1 - \frac{\Gamma_z+\cos(2\varphi)\,\Gamma_\varphi}{\Gamma_+ -
\Gamma_-}\right]    e^{-\Gamma_+\tau } . \quad
    \label{ens-z-sol}\end{eqnarray}
Inserting these results into Eq.\ (\ref{S-Kij-2}) and using relations ${\rm Tr}[\sigma_z \, \rho_{\rm av}(t_1+\tau|1_i)] =z_{\rm av} (t_1+\tau)$ and ${\rm Tr}[\sigma_\varphi \, \rho_{\rm av}(t_1+\tau|1_i)] =z_{\rm av} (t_1+\tau)\cos\varphi +x_{\rm av} (t_1+\tau)\sin\varphi$, we obtain Eqs.\ (14) and (15) of the main text for the correlators $K_{zz}(\tau)$ and $K_{z\varphi}(\tau)$. If $\sigma_{i}\neq\sigma_z$, then we can use rotational symmetry to find $K_{ij}(\tau)$, simply replacing $\varphi$ with the angle difference between the measured directions and renaming the measurement channels.

Note that the evolution (\ref{ens-y-sol}) for the $y$-coordinate was not important for $K_{zz}(\tau)$ and $K_{z\varphi}(\tau)$. Also note that we were able to use Eq.\ (\ref{S-Kij-2}) instead of Eq.\ (\ref{S-Kij}) because for the effective qubit the states $|1\rangle$ and $|0\rangle$ are equivalent (producing unital ensemble-averaged evolution). In similar calculations for a physical (non-rotating) qubit, energy relaxation would make states $|1\rangle$ and $|0\rangle$ non-equivalent, and then we would need to use Eq.\ (\ref{S-Kij}). Finally, we emphasize that this recipe is valid only in the absence of phase backaction. It requires a minor modification when phase backaction is present.

\subsection{Derivation via stochastic Bayesian equations}
\label{sec-FP}

Now let us derive Eqs.\ (14) and (15) of the main text for $K_{zz}(\tau)$ and $K_{z\varphi}(\tau )$ using the
stochastic evolution equations (4)--(6) of the main text instead of the collapse recipe used above. Even though equivalence of these methods was shown in Ref.\ \cite{S-Korotkov2001sp}, we will do the derivation
explicitly, essentially repeating the equivalence proof in
\cite{S-Korotkov2001sp}. In the derivation we assume fixed $t_1$
and $t_1+\tau$ [averaging the correlator (\ref{S-corr-def}) over the ensemble of realizations], and for brevity of notations we assume $t_1=0$. The qubit state
right before the first measurement is therefore $\rho_{\rm
in}\equiv\rho(0)$. (Note that if we have a distribution of the initial states,
it is possible to average the correlator over this distribution
later. However, such averaging is not actually needed because of the linearity of quantum evolution that allows us to use a single initial state, which is equal to the average over the distribution.)

We will mainly consider $K_{z\varphi}(\tau )$; the derivation for $K_{zz}(\tau )$ is similar. Using Eqs.\ (\ref{S-I-z}) and (\ref{S-I-varphi}), we can write the correlator  $K_{z\varphi}(\tau )$ as a sum of two parts, describing a correlation between qubit states at different times and a correlation between the noise
and the future qubit state [there is no correlation with the past states because of causality, and the noise-noise correlations for $\tau > 0$ are also absent for uncorrelated white noises $\xi_z(t)$ and $\xi_\varphi (t)$],
    \begin{eqnarray}
&& K_{z\varphi}(\tau)= K^{(1)}_{z\varphi}(\tau) +
K^{(2)}_{z\varphi}(\tau),
  \label{S-Kzp}\\
&& \hspace{0.0cm} K^{(1)}_{z\varphi}(\tau) = \langle {\rm Tr}[\sigma_\varphi \rho (\tau)] \rangle \, z(0) , \,\,\,
    \label{S-Kzp1}\\
&&  K^{(2)}_{z\varphi}(\tau) = \sqrt{\tau_z}\, \langle {\rm Tr}[\sigma_\varphi \rho (\tau)] \, \xi_z (0)\rangle , \qquad
    \label{S-Kzp2}\end{eqnarray}
where averaging is over the noise realizations $\xi_z(t)$ and $\xi_\varphi
(t)$, which affect evolution of $\rho$ via Eqs. (4)--(6) of the main text, and the initial state is $\rho_{\rm in}=[\openone +x_{\rm in}\sigma_x + y_{\rm in}\sigma_y+z_{\rm in}\sigma_z ]/2$ with $\{ x(0),y(0),z(0)\}=\{x_{\rm in},y_{\rm in},z_{\rm in}\}$.

The first contribution can also be written as
    \be
K^{(1)}_{z\varphi}(\tau) = {\rm Tr}[\sigma_\varphi \rho_{\rm av} (\tau|\rho_{\rm in})] \, z_{\rm in},
    \label{S-Kzp1-2}\ee
where  $\rho_{\rm av} (\tau|\rho_{\rm in})$ is the ensemble-averaged density matrix at time $\tau$, which starts with $\rho_{\rm in}$ at $t=0$. Using linearity of the $\rho_{\rm av}$ evolution given by Eqs.\ (\ref{ens-y-sol}) and (\ref{ens-evol-matrix}), we can {\it formally} rewrite it as
    \be
K^{(1)}_{z\varphi}(\tau) = {\rm Tr}[\sigma_\varphi \rho_{\rm av} (\tau|z_{\rm in}\rho_{\rm in})],
    \label{S-Kzp1-3}\ee
where the evolution of $\rho_{\rm av}$ now starts with state $\rho_{\rm av} (0|z_{\rm in}\rho_{\rm in})=[\openone +z_{\rm in}x_{\rm in}\sigma_x + z_{\rm in}y_{\rm in}\sigma_y+z_{\rm in}^2\sigma_z ]/2$. Note that in the definition of the state $z_{\rm in}\rho_{\rm in}$ we still use physical normalization ${\rm Tr}(z_{\rm in}\rho_{\rm in})=1$, multiplying by $z_{\rm in}$ only Bloch-sphere components of $\rho_{\rm in}$.

To find the second contribution $K^{(2)}_{z\varphi}(\tau)$, we use the
stochastic equations (4)--(6) of the main text [complemented with Eq.~(7) of the main text] and derive the evolution equations for correlators $\langle x(\tau)\,
\xi_z(0)\rangle$ and $\langle z(\tau)\, \xi_z(0)\rangle$:
   \begin{eqnarray}
&& \frac{d}{d\tau} \binom{\langle x(\tau)\, \xi_z(0)\rangle}{\langle
z(\tau)\, \xi_z(0)\rangle} = {\bf M} \binom{\langle x(\tau)\,
\xi_z(0)\rangle}{\langle z(\tau)\, \xi_z(0)\rangle}
    \nonumber \\
&&\hspace{2cm} + \frac{1}{\sqrt{\tau_{z}}} \binom{-x_{\rm
in} z_{\rm in}}{1-z_{\rm in}^2}\, \delta (\tau),
   \label{evol-corr-noise}\end{eqnarray}
where $\bf M$ is the evolution matrix (\ref{M-matrix}), and for $\tau <0$ these correlators are zero because of causality. This equation has a simple physical meaning. As follows from Eqs.\ (4) and (6) of the main text, the noise
$\xi_z(0)$ slightly changes the initial state after an infinitesimal
time $dt$, so that $x(0+dt)=x_{\rm in}-\tau_z^{-1/2}x_{\rm in}z_{\rm
in}\xi_z(0) \, dt$ and $z(0+dt)=z_{\rm in}+\tau_z^{-1/2}(1-z_{\rm
in}^2)\,\xi_z(0) \, dt$. The further evolution starts with this
slightly different state. Therefore, $\langle x(dt)\,
\xi_z(0)\rangle = -\tau_z^{-1/2}x_{\rm in}z_{\rm in}\langle
\xi_{z}(0)^2 dt\rangle =-\tau_z^{-1/2}x_{\rm in}z_{\rm in}$, since $\langle
\xi_{z}(0)^2 dt\rangle = 1$, as follows from Eq.\ (3) of the main text.
Similarly, $\langle z(dt)\, \xi_z(0)\rangle = \tau_z^{-1/2}(1-z_{\rm in}^2)\langle
\xi_{z}(0)^2 dt\rangle =\tau_z^{-1/2}(1-z_{\rm in}^2)$. Thus we obtain the last
term in Eq.\ (\ref{evol-corr-noise}), while the evolution due to the
matrix $\bf M$ is rather obvious. Even though the $y$ component is not
important for our analysis, for generality we can similarly derive $\langle
y(dt)\, \xi_z(0)\rangle = -\tau_z^{-1/2}y_{\rm in}z_{\rm in}\langle
\xi_{z}(0)^2 dt\rangle =-\tau_z^{-1/2} y_{\rm in}z_{\rm in}$.

Since the evolution of  $\langle x(\tau)\, \xi_z(0)\rangle$ and $\langle z(\tau)\, \xi_z(0)\rangle$ is governed by the same matrix $\bf M$ as for the components of $\rho_{\rm av}$ (similar for $y$-component), we can write the contribution $K_{z\varphi}^{(2)}(\tau)$ as
    \begin{eqnarray}
&& K^{(2)}_{z\varphi}(\tau)= {\rm Tr}[\sigma_\varphi \rho_{\rm av}(\tau|\delta \rho_{\rm in})],
    \label{S-Kzp2-res} \\
&& \delta \rho_{\rm in} = \frac{1}{2} [-x_{\rm in}z_{\rm in}\sigma_x
-y_{\rm in} z_{\rm in}\sigma_y+(1-z_{\rm in}^2)\sigma_z ], \qquad
    \end{eqnarray}
where $\delta \rho_{\rm in}$ is an unphysical density matrix with zero trace, in which
the Bloch-sphere components are the shifts discussed above due to the second term in Eq.\ (\ref{evol-corr-noise}), multiplied by $\sqrt{\tau_z}$ because of Eq.\ (\ref{S-Kzp2}).

It is easy to see that
    \be
    z_{\rm in}\rho_{\rm in} + \delta\rho_{\rm in} = [\openone +\sigma_z]/2=|1\rangle \langle 1 |
    \ee
(recall that $z_{\rm in}\rho_{\rm in}$ is defined with unity trace).
Therefore, combining Eqs.\ (\ref{S-Kzp1-3}) and (\ref{S-Kzp2-res}), we find
    \be
    K_{z\varphi}(\tau )= {\rm Tr}[\sigma_\varphi\rho_{\rm av}(\tau|1)],
    \label{S-Kzp-res}\ee
which coincides with Eq.\ (\ref{S-Kij-2}) for $\sigma_i=\sigma_z$ and $\sigma_j=\sigma_\varphi$. Note the slightly different notations for the initial state of $\rho_{\rm av}$, which should not be confusing, for example $\rho_{\rm av}(\tau|1)\equiv \rho_{\rm av}(\tau \big| |1\rangle\langle 1|)$.

Thus we have shown that the correlator $K_{z\varphi}(\tau )$ derived from the stochastic evolution equations coincides with the result of the previous derivation based on the collapse recipe. The derivation for $K_{zz}(\tau)$ from the stochastic equations is similar, we just need to replace $\varphi$ with $z$ and $\sigma_\varphi$ with $\sigma_z$ in Eqs.\ (\ref{S-Kzp})--(\ref{S-Kzp1-3}), (\ref{S-Kzp2-res}), and (\ref{S-Kzp-res}), thus obtaining $K_{zz}(\tau )= {\rm Tr}[\sigma_z \rho_{\rm av}(\tau|1)]$, which coincides with Eq.\ (\ref{S-Kij-2}) for $\sigma_i=\sigma_j=\sigma_z$.

\subsubsection*{Equivalence in a non-unital case}

We have shown equivalence of the results for correlators
$K_{zz}(\tau)$ and $K_{z\varphi}(\tau)$ derived via the stochastic Bayesian equations and via the simple collapse recipe.
However, in showing the equivalence we implicitly used
the fact that the ensemble-averaged equations [Eqs.\ (11)-(13) of the main text] are homogeneous (not only linear). This is the so-called unital evolution (which preserves the center of the Bloch sphere), which originates from full symmetry between the states $|0\rangle$ and $|1\rangle$ of the effective qubit. Let us now prove that even in a non-unital case (for example, when we measure a physical qubit and asymmetry between states $|0\rangle$ and $|1\rangle$ is created by energy relaxation), the two methods for calculation of correlators are still equivalent (in the absence of phase-back-action). We will see that in this general case we can use Eq.\ (\ref{S-Kij}) originating from the collapse recipe, but cannot use its simplified version (\ref{S-Kij-2}).

Let us use the linearity of the ensemble-averaged quantum evolution $\cal E$ (a trace-preserving positive map) from  $t=0$ to $t=\tau$,
    \be
    \rho_{\rm in} \rightarrow {\cal E}(\rho_{\rm in})=\tilde\rho_c + x_{\rm in} \Delta\tilde\rho_x +  y_{\rm in} \Delta\tilde\rho_y +  z_{\rm in} \Delta\tilde\rho_z,
    \label{general-evol}\ee
where $\tilde\rho_c ={\cal E}(\rho_c)$ is the state mapped from the Bloch sphere center $\rho_c=\openone /2$, while $\Delta\tilde\rho_x ={\cal E}(\rho_x)-\tilde\rho_c$, $\Delta\tilde\rho_y={\cal E}(\rho_y)-\tilde\rho_c$, and $\Delta\tilde\rho_z={\cal E}(\rho_z) -\tilde\rho_c$ describe mapping of the Bloch sphere axes, with density matrices $\rho_x$, $\rho_y$, and $\rho_z$ corresponding to pure states $(|1\rangle +|0\rangle)/\sqrt{2}$, $(|1\rangle +i |0\rangle)/\sqrt{2}$, and $|1\rangle$, respectively. Following the same logic as above, we can write the first contribution (\ref{S-Kzp1-2}) to $K_{z\varphi}(\tau)$ as
    \begin{eqnarray}
&&  K_{z\varphi}^{(1)}(\tau) =
    [{\rm Tr}(\sigma_\varphi \tilde\rho_c) +x_{\rm in} {\rm Tr}(\sigma_\varphi \Delta\tilde\rho_x) +y_{\rm in}{\rm Tr}(\sigma_\varphi \Delta\tilde\rho_y)
    \nonumber \\
&& \hspace{1.7cm}    +z_{\rm in}{\rm Tr}(\sigma_\varphi \Delta\tilde\rho_z)]\, z_{\rm in}.
    \end{eqnarray}
The second contribution (\ref{S-Kzp2-res}) to $K_{z\varphi}(\tau)$ can be written as
   \begin{eqnarray}
&& K_{z\varphi}^{(2)}(\tau) =
       -x_{\rm in}z_{\rm in} {\rm Tr}(\sigma_\varphi \Delta\tilde\rho_x)   -y_{\rm in}z_{\rm in} {\rm Tr}(\sigma_\varphi \Delta\tilde\rho_y)
       \nonumber \\
&& \hspace{1.7cm}  +(1-z_{\rm in}^2) \, {\rm Tr}(\sigma_\varphi \Delta\tilde\rho_z).
          \end{eqnarray}
Combining the two contributions, we find
    \be
      K_{z\varphi}(\tau) =
        {\rm Tr}(\sigma_\varphi \tilde\rho_c) \, z_{\rm in} + {\rm Tr}(\sigma_\varphi \Delta\tilde\rho_z).
    \label{K-zphi-der2-f1}\ee

On the other hand, using Eq.\ (\ref{S-Kij}) of the collapse recipe with $\sigma_i=\sigma_z$ and $\sigma_j=\sigma_\varphi$ , we obtain
    \begin{eqnarray}
&& K_{z\varphi}(\tau) = [ {\rm Tr}(\sigma_\varphi \tilde\rho_c)   + {\rm Tr}(\sigma_\varphi \Delta\tilde\rho_z)]\, \frac{1+z_{\rm in}}{2}
 \nonumber \\
&& \hspace{1.4cm} -  [{\rm Tr}(\sigma_\varphi \tilde\rho_c)  - {\rm Tr}(\sigma_\varphi \Delta\tilde\rho_z)]\, \frac{1-z_{\rm in}}{2},
    \end{eqnarray}
which coincides with Eq.\ (\ref{K-zphi-der2-f1}). Thus, equivalence of both methods for $K_{z\varphi}(\tau)$ is proven in the general (non-unital) case. The proof for the correlator  $K_{zz}(\tau)$ is practically the same, just replacing $\sigma_\varphi$ with $\sigma_z$. The proof of the equivalence for an arbitrary $K_{ij}(\tau)$ is also similar, but we need to use the basis, corresponding to $\sigma_i$.

Note that in the non-unital case we should use Eq.\ (\ref{S-Kij}) of the collapse recipe, which takes into account both scenarios (collapse to the state $|0\rangle$ or to $|1\rangle$) and not the simplified equation (\ref{S-Kij-2}) (collapse to $|1\rangle$ only), which is valid only for a symmetric (unital) evolution. As seen from Eq.\ (\ref{S-Kij}) [or Eq.\ (\ref{K-zphi-der2-f1})], in the non-unital case the correlator $K_{ij}(\tau)$ {\it depends} on the initial state $\rho_{\rm in}$ via the term ${\rm Tr} [\sigma_j \rho_{\rm av} (t_1+\tau |\rho_{\rm c})]\, {\rm Tr} [\sigma_i \rho (t_1)]$, where $\rho_{\rm c}$ is the fully mixed state.

Also note that in our derivation we assumed absence of the phase backaction terms \cite{S-Korotkov-99-01, S-Korotkov-2011, S-Korotkov-2016, S-Gambetta2008} in the Bayesian stochastic equations. These terms would introduce additional contribution to $\delta\rho_{\rm in}$ and therefore to correlators. The collapse recipe in this case should be modified accordingly.

\subsection{Derivation via quantum regression approach}
\label{sec-regression}

Now let us derive Eqs.\ (14) and (15) of the main text for the
correlators $K_{zz}(\tau)$ and $K_{z\varphi}(\tau)$ using the
standard non-stochastic approach \cite{S-Gardiner-book}, which cannot describe
individual realizations of the qubit measurement process, but is
sufficient to calculate correlators. In this section we
assume $t_1=0$ and use $\rho_{\rm in}=\rho(0)$.

In this approach \cite{S-Gardiner-book} we need to use the Heisenberg picture and associate
the measurement outcomes $I_z(t)$ and $I_\varphi (t)$ with the
operators $\sigma_z(t)$ and $\sigma_\varphi (t)$, which evolve in
time as $\sigma_z (\tau) \equiv e^{i H_{\rm tot}\tau} \sigma_z
e^{-iH_{\rm tot}\tau}$ and $\sigma_\varphi (\tau) \equiv e^{i H_{\rm
tot}\tau} \sigma_\varphi e^{-iH_{\rm tot}\tau}$, where $H_{\rm tot}$
is the total Hamiltonian describing the qubit, environment, and
interaction between them (in this approach we consider the measurement
apparatus as an environment). The correlators of the outcomes then can
be expressed as symmetrized combinations
    \begin{eqnarray}
    && K_{zz}(\tau) = \frac{1}{2} {\rm Tr_{tot}} [
\sigma_z(\tau)\, \sigma_z(0)\, \rho_{\rm tot}(0)
    \nonumber \\
&&\hspace{2.4cm}
 + \sigma_z(0)\,
\sigma_z(\tau) \, \rho_{\rm tot}(0)]
    \nonumber\\
&& \hspace{1.2cm} = {\rm Re} \{ {\rm Tr_{tot}} [ \sigma_z (\tau) \,
\sigma_z(0) \, \rho_{\rm tot}(0) ]\},
    \\
 && K_{z\varphi}(\tau) = \frac{1}{2} {\rm Tr_{tot}}[ \sigma_\varphi(\tau)\, \sigma_z(0)\, \rho_{\rm tot}(0)
    \nonumber \\
 && \hspace{2.4cm} + \sigma_z(0)\, \sigma_\varphi(\tau)
\, \rho_{\rm tot}(0)]
    \nonumber\\
&& \hspace{1.2cm} = {\rm Re} \{ {\rm Tr_{tot}} [ \sigma_\varphi
(\tau) \, \sigma_z(0) \, \rho_{\rm tot}(0) ]\},
    \end{eqnarray}
where $\rho_{\rm tot}(0)=\rho (0) \otimes \rho_{\rm env}(0)$ is the
initial density matrix, which includes the environment (``bath''),
and the trace should be taken over the qubit and environment degrees
of freedom.

We need to assume that the coupling between the qubit and the
environment is sufficiently weak, so that the effective decoherence
rate of the qubit due to its coupling with the environment is much
smaller than the reciprocal of the typical correlation time for the
bath degrees of freedom (this includes the ``bad cavity'' assumption). In this case we can use the standard formula \cite{S-Gardiner-book}
(related to what is usually called the Quantum Regression Theorem)
    \be
\label{QRT} {\rm Tr_{\rm tot}} [ A (\tau)\, B(0) \,\rho_{\rm tot}
(0)] = {\rm Tr }_{\rm sys} [ A \,\rho_{\rm av}(\tau|B\rho_{\rm
in})] ,
    \ee
where in the right-hand side the trace is only over the system (qubit),
operators $A$ and $B$ are system observables, and $\rho_{\rm av}(\tau|B\rho_{\rm in})$ is the system (reduced) density matrix at
time $\tau$, which evolves in time according to the
ensemble-averaged (reduced) evolution equations and starts in the
state $B\rho_{\rm in}$, i.e., $\rho_{\rm av}(0|B\rho_{\rm in})=B\rho_{\rm in}$.
Note that $\rho_{\rm av}(\tau|B\rho_{\rm in})$ is unphysical
because it starts with an unphysical initial state $B\rho_{\rm in}$ (it is typically not Hermitian and not normalized). Also note that the
validity of Eq.~\eqref{QRT} requires that the system and the
environment will be weakly entangled; i.e., $\rho_{\rm tot}(t)\approx
{\rm Tr}_{\rm env}[\rho_{\rm tot}(t)]\otimes \rho_{\rm env}(0)$.
This is consistent with the above assumption that the coupling is
weak.

In our case in Eq.\ (\ref{QRT}) the operator $B$ is $\sigma_z$,
while $A$ is either $\sigma_z$ or $\sigma_\varphi$. The starting
state for $\rho_{\rm av}(\tau|B\rho_{\rm in})$ is $\sigma_z\rho_{\rm in}$, so
    \begin{eqnarray}
    &&  K_{zz}(\tau) = {\rm Re} \{ {\rm Tr}
    [ \sigma_z   \rho_{\rm av}(\tau|\sigma_z \rho_{\rm in}) ]\},
    \label{K-zz-der3}\\
   &&  K_{z\varphi}(\tau) = {\rm Re} \{ {\rm Tr}
    [ \sigma_\varphi   \rho_{\rm av}(\tau|\sigma_z \rho_{\rm in}) ]\}.
    \label{K-zphi-der3}\end{eqnarray}
Since we have to work with unphysical unnormalized states, we use $\rho(t) = [P_{\rm N} \openone
+ x(t)\, \sigma_x + y(t)\, \sigma_y + z(t)\, \sigma_z]/2$,
where the normalization is conserved, $\dot{P}_{\rm N}=0$, during the ensemble-averaged evolution described by Eqs.\ (11)--(13) of the main text.

Now let us represent the unphysical initial state $\sigma_z\rho_{\rm in}$ as
    \begin{eqnarray}
&& \sigma_z \begin{bmatrix} \rho_{11,\rm in} &
   \rho_{10, \rm in}
   \\ (\rho_{10, \rm in})^*
   & \rho_{00, \rm in} \end{bmatrix} =
   \frac{\sigma_z}{2}+(\rho_{11,\rm in}-\rho_{00,\rm in})\frac{\openone}{2}
   \nonumber \\
&& \hspace{3.3cm}
+    \begin{bmatrix} 0 &
   \rho_{10,\rm in}
   \\ -(\rho_{10,\rm in})^*
   & 0 \end{bmatrix} ,
    \label{unphys-in}\end{eqnarray}
using condition $\rho_{00,\rm in}+\rho_{11,\rm in}=1$. Since the
ensemble-averaged evolution is linear,
we can separate $\rho_{\rm av}(\tau|\sigma_z\rho_{\rm in})$ into
three terms, corresponding to the three terms in Eq.\
(\ref{unphys-in}). The first term, $\sigma_z/2$, gives the physical evolution $\rho_{\rm av}(\tau |1)$ starting with the state $|1\rangle$. The second term, $(\rho_{11,\rm in} -\rho_{00,\rm in})\,\openone /2$, does not change in time and gives zero contribution to the correlators (\ref{K-zz-der3}) and (\ref{K-zphi-der3}). The third term is initially anti-Hermitian, and it will remain anti-Hermitian
in the evolution, because all coefficients in Eqs.\ (11)--(13) of the main text are real. The anti-Hermitian term will
give zero contribution to Eqs.\ (\ref{K-zz-der3}) and (\ref{K-zphi-der3}) because the traces will be
imaginary numbers.

Thus we obtain equations
    \be
K_{zz}(\tau) = {\rm Tr}
    [ \sigma_z   \rho_{\rm av}(\tau|1)], \,\,\,
    K_{z\varphi}(\tau) = {\rm Tr}
    [ \sigma_\varphi   \rho_{\rm av}(\tau|1)],
    \ee
which coincide with Eq.\ (\ref{S-Kij-2}) for $\sigma_i=\sigma_z$. Therefore, the final result for the correlators is the same as for the derivation based on the collapse recipe.

In the general non-unital case (without phase back-action), using representation (\ref{general-evol}) of a linear quantum map, we can obtain Eq.\ (\ref{K-zphi-der2-f1}) from Eqs.\ (\ref{K-zphi-der3}) and (\ref{unphys-in}), thus proving that the derivation via the quantum regression approach is still equivalent to the derivations via the stochastic equations and via the collapse recipe.

\section{Extracting correlators from experimental data}

The experimental correlators are calculated as
\begin{align}
K_{ij} (\tau) = \int_{t_a}^{t_b}dt_1\, \frac{\left\langle \big(\tilde I_i(t_1)-\tilde I_i^{\rm off}\big)\big(\tilde I_j(t_1+\tau)-\tilde I_j^{\rm off}\big)\right\rangle}{(t_b-t_a) \,\Delta \tilde I_i \, \Delta \tilde I_j},
\label{eq:Kij-expt}
\end{align}
where $\tilde I_i(t)$ are experimental output signals for $\sigma_i$ measurement ($0\leq t\leq 5\,\mu$s), $\langle ...\rangle$ denotes ensemble averaging over all selected traces with the same angle difference $\varphi$ ($\sim$200,000 per angle, the selection includes heralding at the start of the run and checking that the physical transmon qubit is in the subspace of its lowest two energy levels after the run), additional time-averaging is between $t_a=1\,\mu$s and $t_b=1.5\,\mu$s, the correlators are normalized by responses $\Delta \tilde{I}_i$, and small offsets $\tilde I_i^{\rm off}$ are calculated separately for each value of $\varphi$ (see below). Significantly larger offsets are already removed from $\tilde I_i(t)$ individually for each trace by measuring and averaging the background noise for the non-rotating qubit after each trace.

\begin{figure}[t!]
\centering
\includegraphics[width=8cm, trim = 4.0cm 3.0cm 3.0cm 3.0cm,clip=true]{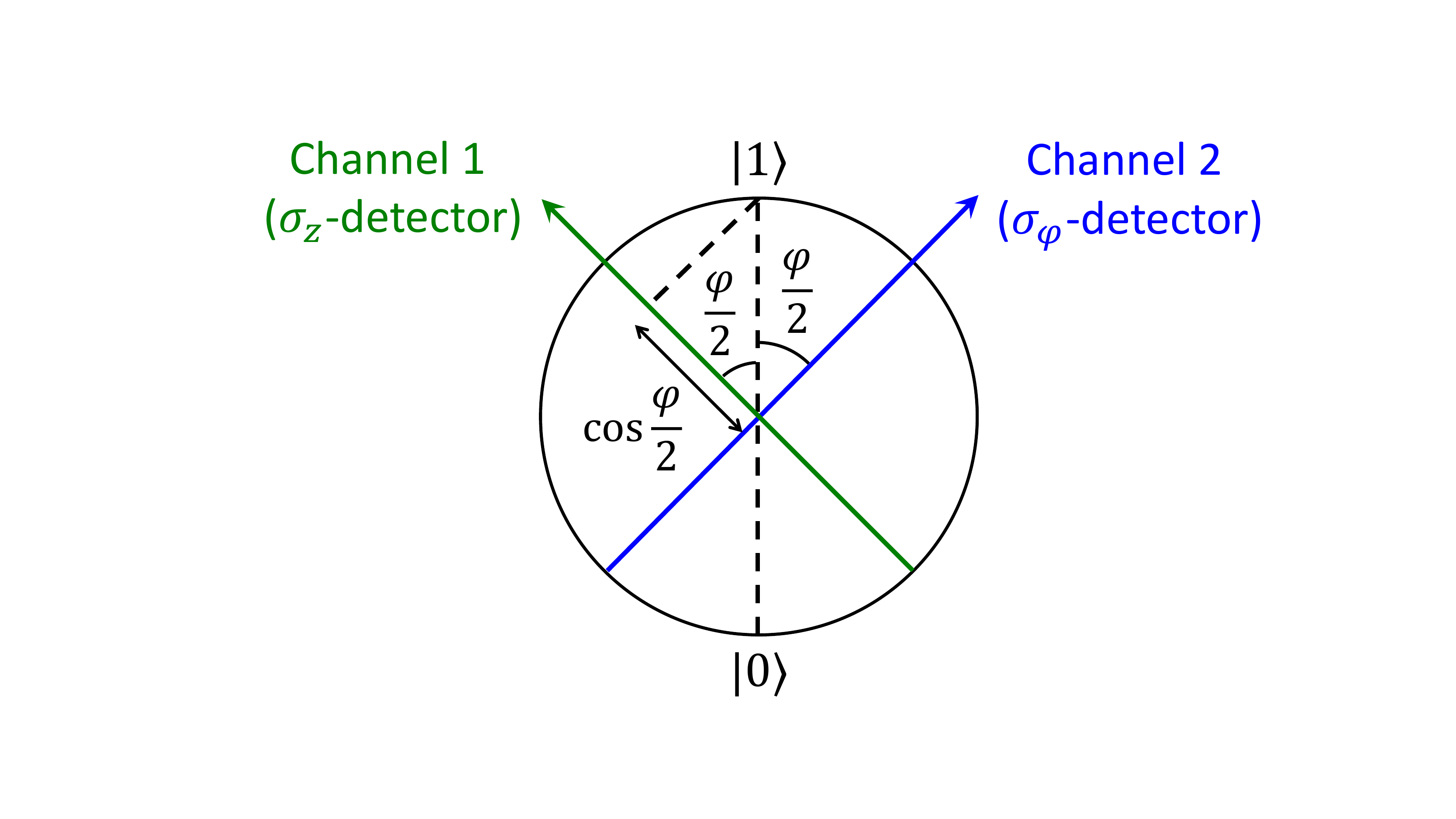}
\caption{Bloch $xz$-plane of the effective qubit. The $z$-axis is calibrated for $\varphi =0$ (nominally, neglecting small $\delta \varphi$), while for non-zero angle difference $\varphi$, the stroboscopic measurement axes are nominally at $-\varphi/2$ (channel 1) and $\varphi/2$ (channel 2). Then for the effective qubit with $z=1$, the average signals for both channels are proportional to $\cos (\varphi/2)$.
 }
\label{fig:effective-qubit-xz-plane}
\end{figure}

To find $\Delta \tilde I_i$ and $\tilde I_i^{\rm off}$, for each angle $\varphi$ we separate the traces (each trace includes outputs for both measurement channels) into two approximately equal groups. These groups correspond to the effective qubit initialized either in the state $|1\rangle$ ($z_0=1$) or $|0\rangle$ ($z_0=-1$), which is controlled by the initial state $|1\rangle$ or $|0\rangle$ of the physical qubit before application of 40 MHz Rabi oscillations and stroboscopic sideband measurement. Calibration of the $z$-axis of the effective qubit is done by maximizing the response (for the lower-$\kappa$ channel) for zero nominal angle between the two measurement directions ($\varphi=0$ neglecting small $\delta \varphi$). For non-zero nominal $\varphi$, the stroboscopic measurement directions are $-\varphi/2$ (channel 1, $\omega_{\rm r,1}/2\pi=7.4\, {\rm GHz}$, $\kappa_1/2\pi=4.3\,{\rm MHz}$) and $\varphi/2$ (channel 2, $\omega_{\rm r,2}/2\pi=6.7\, {\rm GHz}$, $\kappa_2/2\pi=7.2\,{\rm MHz}$) -- see Fig.~\ref{fig:effective-qubit-xz-plane}. Theoretically only the angle difference $\varphi$ matters for the correlators; however, for calibration we need to use the actual measurement directions $\pm \varphi/2$ (more accurately, $-\varphi/2$ and $\varphi/2+\delta\varphi$). In the main text the channel 1 is called $z$-channel, while channel 2 is $\varphi$-channel. In this section we will also be using the terminology of channels 1 and 2.

\begin{figure}[t!]
\centering
\begin{tabular}{cc}
\includegraphics[width=8.5cm, trim = 2.8cm 1.5cm 3.5cm 2cm,clip=true]{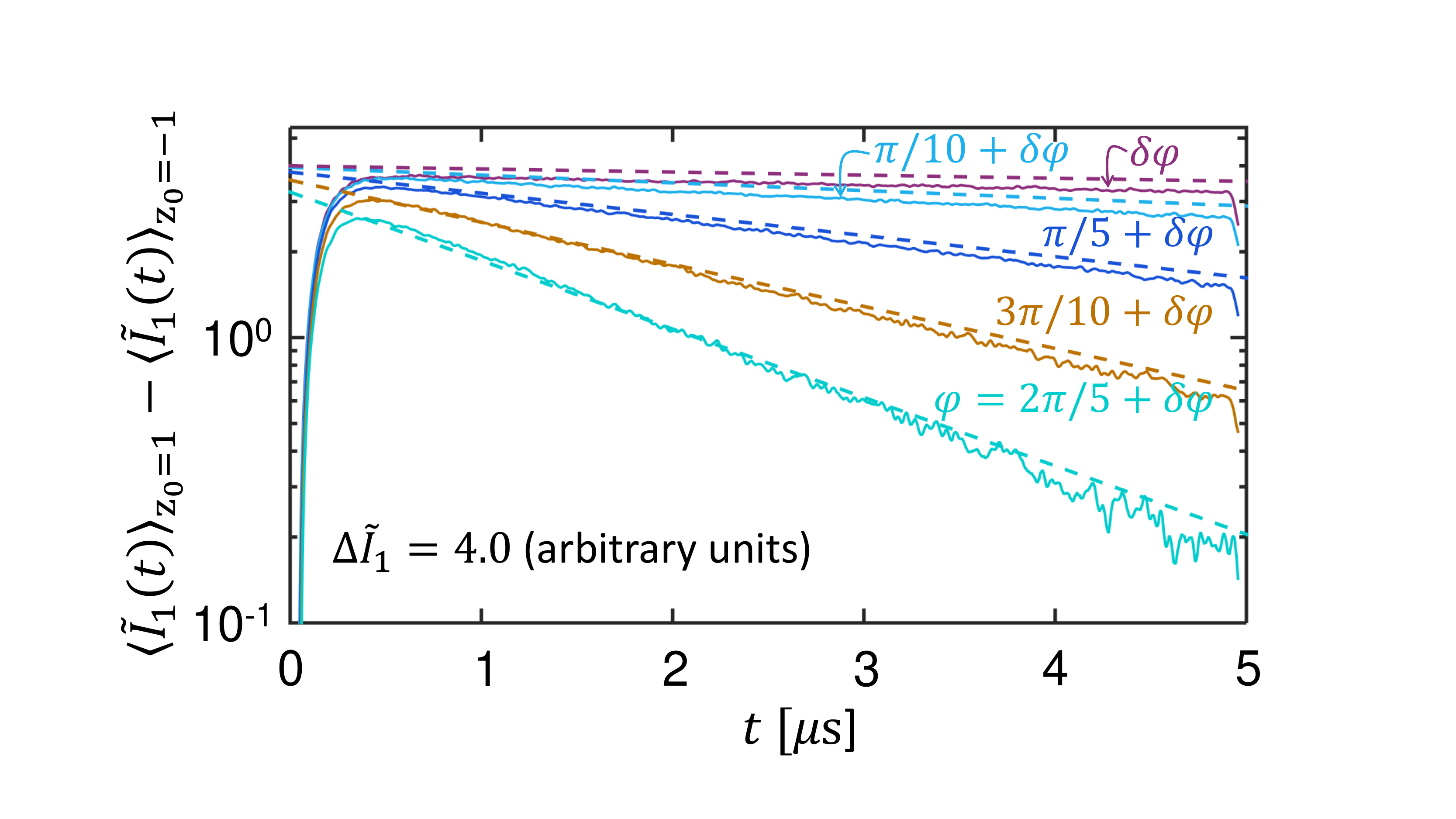}
\\
\includegraphics[width=8.5cm, trim=2.8cm  1.5cm 3.5cm 2cm,clip=true]{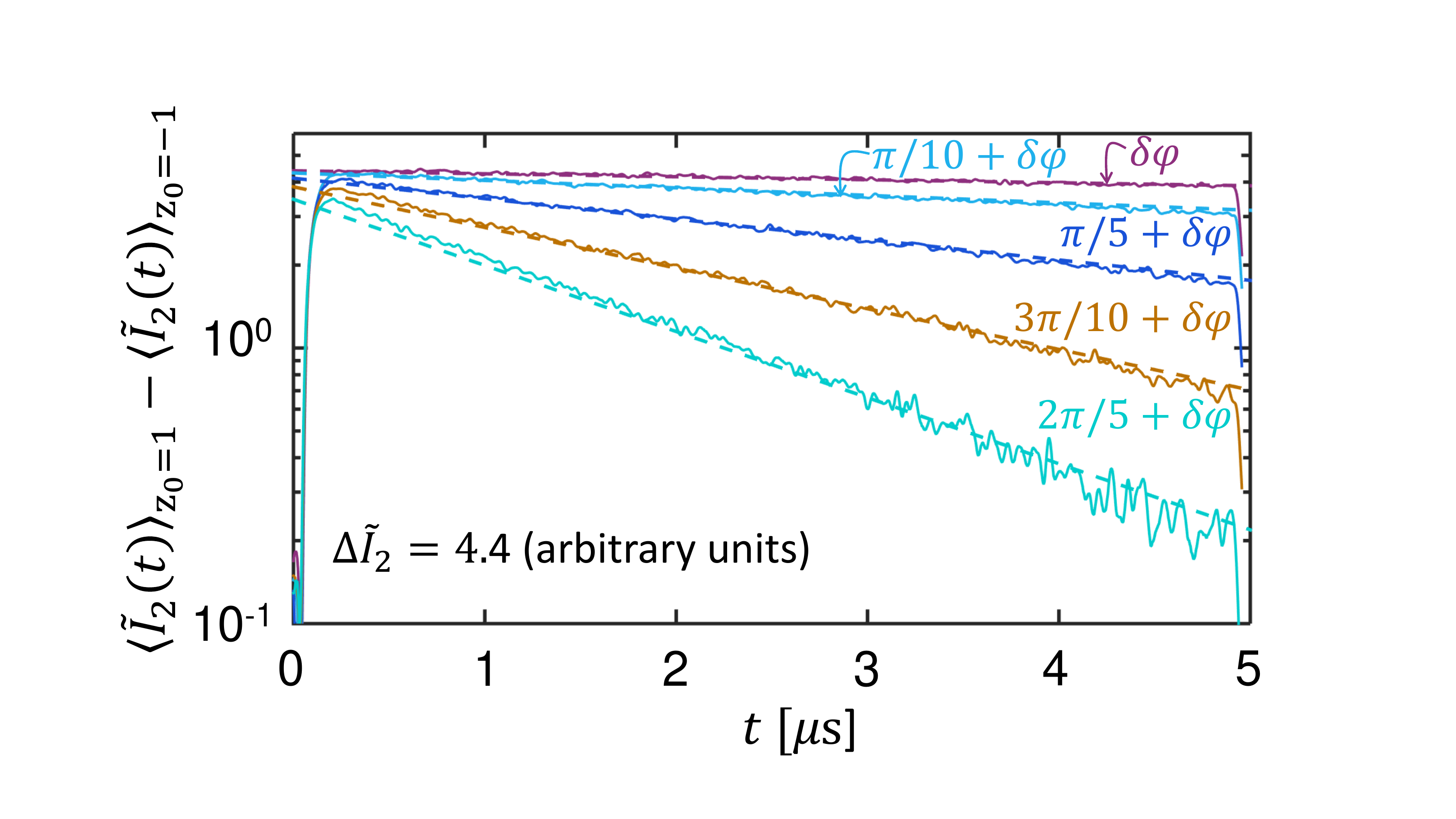}
\end{tabular}
\caption{Finding detector responses $\Delta \tilde{I}_i$ from experimental data for channel 1 (upper panel) and channel 2 (lower panel). Solid lines show experimental data for the difference $\langle \tilde{I}_i(t)\rangle_{z_0=1}-\langle \tilde{I}_i(t)\rangle_{z_0=-1}$  between ensemble-averaged output signals with the effective qubit initialized either at $z_0=1$ or $z_0=-1$ at $t=0$, for 5 values of the angle difference $\varphi$. The dashed lines are analytical results obtained from the ensemble-averaged evolution with fitted response values $\Delta \tilde I_1\equiv \Delta \tilde I_z  = 4.0$ and $\Delta \tilde I_2 \equiv \Delta \tilde I_\varphi = 4.4$.
Note that the dashed lines do not cross at $t=0$ since the theoretical value (neglecting $\delta\varphi$) is $\Delta \tilde I_i\cos(\varphi/2)$.
}
\label{fig:Finding-responses}
\end{figure}

To find the responses $\Delta \tilde I_1\equiv\Delta \tilde I_z$ and  $\Delta \tilde I_2\equiv\Delta \tilde I_\varphi$ for the two channels, we calculate $\mathcal{D}_i(t)\equiv\langle \tilde I_i(t)\rangle_{z_0=1} - \langle \tilde I_i(t)\rangle_{z_0=-1}$ for each $\varphi$, where the subscripts $z_0=\pm 1$ denote the group of traces with initial state $z_0$ of the effective qubit. This quantity can be also obtained theoretically from the ensemble averaged evolution equations (11)--(13) of the main text with the initial condition $x(0)=\sin(\varphi/2)$, $y(0)=0$, and $z(0)=\cos(\varphi/2)$, see Fig.~\ref{fig:effective-qubit-xz-plane}. It is equal to $\mathcal{D}_1(t)= \Delta \tilde I_1 z_{\rm av}(t)$ and $\mathcal{D}_2(t) = \Delta \tilde I_2 [z_{\rm av}(t)\cos(\varphi+\delta\varphi) + x_{\rm av}(t)\sin(\varphi+\delta\varphi)]$ for the first and second channels, respectively. In Fig.~\ref{fig:Finding-responses}, we plot experimental $\mathcal{D}_1(t)$ and $\mathcal{D}_2(t)$ for 5 values of $\varphi$ ($\varphi_n=n\pi/10$, $n=0,1,2,3,4$), and fit them with theoretical results. We find a good agreement for the responses $\Delta \tilde I_1 = 4.0$ and $\Delta \tilde I_2 = 4.4$ in units of the experimental output. In this fitting we disregard any residual Rabi oscillations ($\tilde \Omega_{\rm R}=0$).

\begin{figure}[t]
\centering
\includegraphics[width=8.5cm, trim = 3cm 1.2cm 4cm 1.2cm,clip=true]{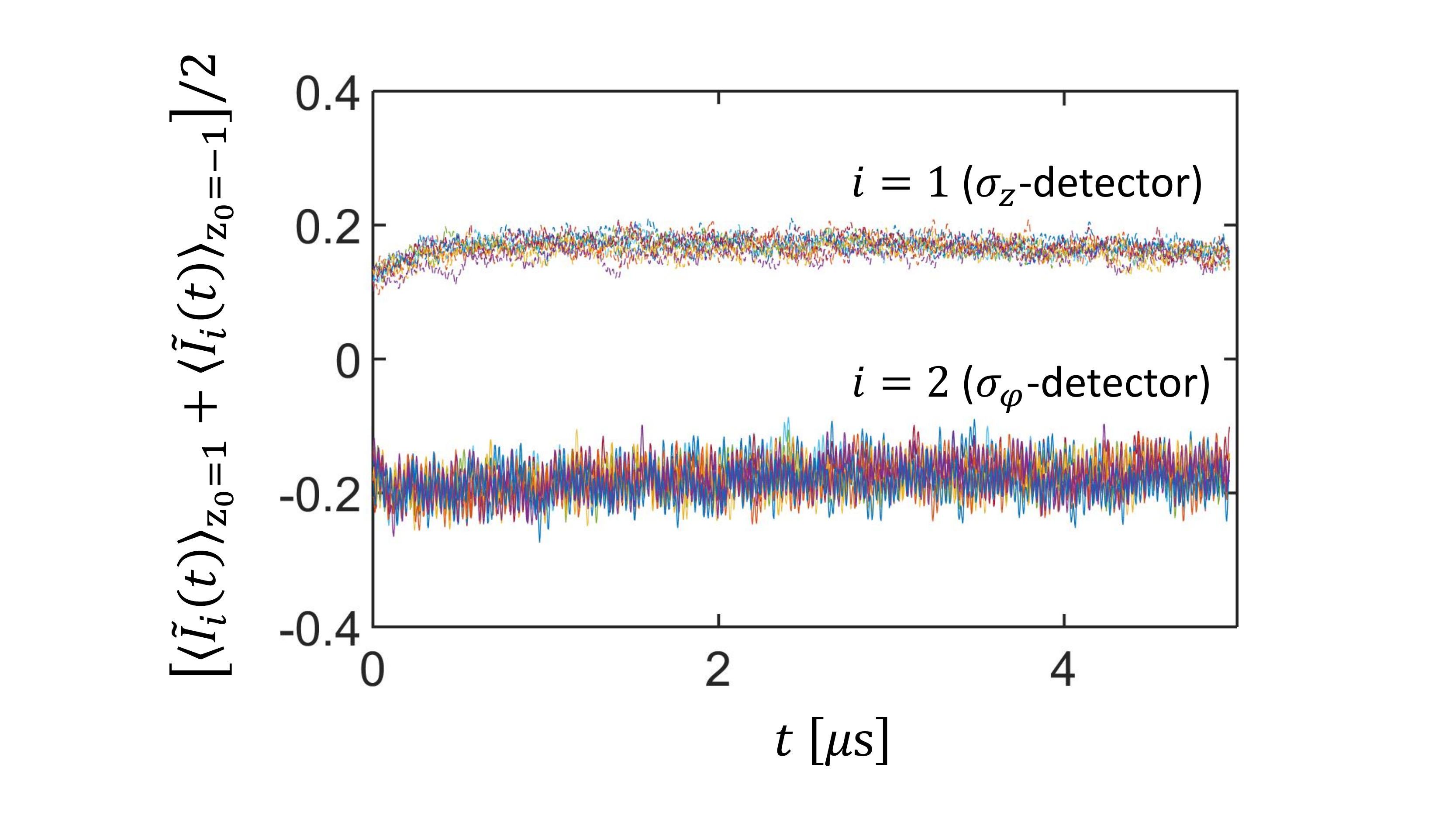}
\caption{Finding the offsets $\tilde I_i^{\rm off}$ from experimental data, using the symmetric combination $[\langle \tilde{I}_i(t) \rangle_{z_0=1} +\langle \tilde{I}_i(t) \rangle_{z_0=-1}]/2$.
The dashed lines with values close to $0.2$ correspond to the channel 1 ($\sigma_z$-detector) for 11 values of the angle $\varphi$ (different colors). Similarly, the solid lines with values close to $-0.2$ correspond to the channel 2 ($\sigma_\varphi$-detector). }
\label{fig:offsets}
\end{figure}

To estimate the offsets $\tilde I_i^{\rm off}$, we use the symmetric combination $\mathcal{S}(t) \equiv [\langle \tilde I_i(t)\rangle_{z_0=1} + \langle \tilde I_i(t)\rangle_{z_0=-1}]/2$, which is shown in Fig.~\ref{fig:offsets} for both measurement channels and for 11 angles $\varphi$. We see that the $\mathcal{S}(t)$ are approximately independent of time, and therefore we can introduce the offsets $\tilde I_i^{\rm off}=\mathcal{S}$ for each channel and each value of $\varphi$. The offsets only weakly depend on the angle $\varphi$ , but are significantly different in the two channels. For the first ($z$) channel we crudely find $\tilde I_1^{\rm off}\equiv \tilde I_z^{\rm off}  = 0.15$, 0.16, 0.16, 0.16, 0.17, 0.16, 0.16, 0.17, 0.16, 0.17 0.18 for the angles $\varphi$ in increasing order. For the second ($\varphi$) channel we find $\tilde I_2^{\rm off}\equiv \tilde I_\varphi^{\rm off} = -0.18$, $-0.17$, $-0.17$, $-0.15$, $-0.16$, $-0.17$, $-0.17$, $-0.17$, $-0.18$, $-0.19$ and $-0.19$ for the angles $\varphi$ in increasing order.

\section{Self-correlators at small $\tau$}

In this section we discuss why the amplified vacuum noise is still white (delta-correlated) for finite damping rate $\kappa$ of a resonator. We also estimate the self-correlator contribution $K_{ii}(\tau) \propto \exp (-\kappa_i \tau/2)$, with small amplitude $\sim \Gamma /\kappa_i$ due to qubit evolution.

\begin{figure}[t]
\centering
\includegraphics[width=8.5cm, trim = 9.5cm 7.5cm 10cm 8.5cm,clip=true]{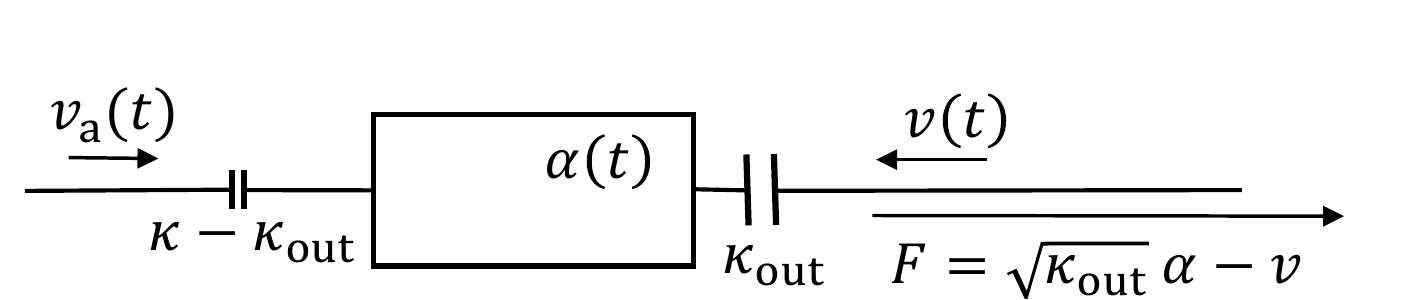}
\caption{Schematic of a resonator with damping rate $\kappa$, coupled with the output transmission line with strength $\kappa_{\rm out}$, while the remaining dissipation rate $\kappa-\kappa_{\rm out}$ is ascribed to another transmission line. The vacuum noises $v(t)$ and $v_{\rm a}(t)$ are incident to the resonator from the transmission lines; they create fluctuating resonator field $\alpha(t)$. }
\label{fig-res-kappa}
\end{figure}

It is a somewhat surprising result that the self-correlator $K_{ii}(\tau)$ does not have a significant contribution $\propto \exp (-\kappa_i \tau/2)$, originating from the correlation time $(\kappa_{i}/2)^{-1}$ of vacuum fluctuations inside the resonator. [We may naively expect that this would widen the contribution $K_{ii} (\tau)=\eta_i\tau_i\delta(\tau)$ from the amplified vacuum noise.] To show why this is not the case, let us consider a resonator with finite $\kappa$ without a qubit (Fig.\ \ref{fig-res-kappa}) and calculate the correlator for the amplified vacuum noise, coming from the resonator. The coupling between the resonator and the output transmission line is $\kappa_{\rm out}$, while the remaining dissipation rate $\kappa-\kappa_{\rm out}$ is modeled as a coupling to another transmission line.

Using the standard input-output theory \cite{S-Gardiner-book, S-Gardiner-1985, S-Clerk-2010}, we need to consider the vacuum noise $\hat{v}(t)$ incident to the resonator from the output line, with the operator correlator $\langle \hat{v}(t)\,\hat{v}^\dagger(t')\rangle=\delta (t-t')$, and write the equations for the annihilation operators in the Heisenberg picture. However, for our purposes it is sufficient to use a simpler approach (e.g., Appendix B of \cite{S-Korotkov-2016}), in which we consider the evolution of classical fields (in the usual, i.e., Schr\"odinger picture) due to ``classical'' vacuum noise $v(t)$ (complex number) with the correlator
         \be
    \langle v(t)\, v^*(t')\rangle = \frac{1}{2}\, \delta (t-t'), \,\,\, \langle v(t)\, v(t')\rangle =0,
    \label{S-vacuum-corr}\ee
where real and imaginary parts of $v(t)$ correspond to orthogonal quadratures. As follows from Eq.\ (\ref{S-vacuum-corr}), any quadrature has correlator $(1/4)\,\delta (t-t')$, and orthogonal quadratures are uncorrelated.

The evolution of the resonator field fluctuation $\alpha(t)$ (evolution of the field due to drive is decoupled due to linearity) is
    \be
    \dot{\alpha}=-i \Delta\omega \, \alpha -\frac{\kappa}{2}\, \alpha
    +\sqrt{\kappa_{\rm out}}\, v(t) +\sqrt{\kappa -\kappa_{\rm out}}\, v_{\rm a}(t),
    \label{S-alpha-dot}\ee
where $\Delta \omega=\omega_{\rm r}-\omega_{\rm d}$ is the resonator frequency $\omega_{\rm r}$ in the rotating frame based on the drive frequency $\omega_{\rm d}$ (it is important for homodyne detection) and $v_{\rm a}(t)$ is the additional vacuum noise from the other transmission line (see Fig.\ \ref{fig-res-kappa}) with the same correlator (\ref{S-vacuum-corr}) and uncorrelated with $v(t)$. In Eq.\ (\ref{S-alpha-dot}) we use the standard normalizations for the resonator field (based on the average number of photons) and for the propagating fields (based on average number of propagating photons per second). This equation has the simple solution,
    \be
    \alpha (t) = \int_{-\infty}^{t} [\sqrt{\kappa_{\rm out}}\, v(t')+
    \sqrt{\kappa -\kappa_{\rm out}}    \, v_{\rm a}(t')] \, e^{-\tilde{\kappa}(t-t')/2}\, dt',
    \label{S-alpha-sol}\ee
where $\tilde\kappa = \kappa +2i\Delta\omega$.

The outgoing field $F(t)$, which is then amplified is
    \be
    F(t) = -v(t)+\sqrt{\kappa_{\rm out}}\, \alpha (t).
    \label{S-F-def}\ee
After a phase-sensitive amplification, this field is sent to a homodyne detector, which outputs the signal $I(t)\propto {\rm Re}[F(t)\,e^{-i\theta}]$, where $\theta$ is the amplified quadrature. Without loss of generality, we can assume $\theta=0$ (by properly defining the quadrature). Therefore, we are interested in the self-correlator of ${\rm Re}[F(t)]$, which is equal (up to a coefficient) to the output signal correlator.

Our goal is to show that the correlator of ${\rm Re}[F(t)]$ is the same as for vacuum noise, i.e.,
    \be
    K_{{\rm Re}F}(\tau)\equiv \langle {\rm Re}[F(t)] \, {\rm Re}[F(t+\tau)]\rangle = \frac{1}{4} \, \delta(\tau ).
    \label{S-K-F}\ee
Actually, it is sufficient to show that $K_{{\rm Re}F}(\tau )=0$ for $\tau >0$, since the coefficient $1/4$ in Eq.\ (\ref{S-K-F}) can be simply obtained from Eqs.\ (\ref{S-vacuum-corr}) and (\ref{S-F-def}). It is also sufficient to choose $t=0$ in Eq.\ (\ref{S-K-F}).

As follows from Eq.\ (\ref{S-F-def}), the correlator $K_{{\rm Re}F}(\tau)$ for $\tau\geq 0$ has three contributions, $K_{{\rm Re}F}(\tau)=K_{{\rm Re}F}^{(1)}(\tau) +K_{{\rm Re}F}^{(2)}(\tau)+K_{{\rm Re}F}^{(3)}(\tau)$, where
    \begin{eqnarray}
&& K_{{\rm Re}F}^{(1)}(\tau) = \langle {\rm Re}[v(0)] \, {\rm Re}[v(\tau)]\rangle  = \frac{1}{4}\, \delta(\tau), \quad
    \label{S-KF1}\\
&& K_{{\rm Re}F}^{(2)}(\tau) = -\sqrt{\kappa_{\rm out}}\, \langle {\rm Re}[v(0)] \, {\rm Re}[\alpha(\tau)]\rangle  \quad
    \label{S-KF2}\\
&& K_{{\rm Re}F}^{(3)}(\tau) =\kappa_{\rm out}\langle {\rm Re}[\alpha(0)] \, {\rm Re}[\alpha(\tau)]\rangle .  \quad
    \label{S-KF3}\end{eqnarray}
Note that for $\tau \geq 0$ there is no contribution due to correlation between $\alpha(0)$ and $v (\tau)$ because of causality. We need to show that the second and third contributions exactly cancel each other, $K_{{\rm Re}F}^{(2)}(\tau)+K_{{\rm Re}F}^{(3)}(\tau)=0$.

Using Eq.\ (\ref{S-alpha-sol}), correlator $\langle {\rm Re}[v(t)]\, {\rm Re}[v(t')]\rangle =(1/4)\,\delta (t-t')$, and absence of correlation between ${\rm Re}[v(t)]$ and ${\rm Im}[v(t')]$, we easily obtain ($\tau > 0$)
        \be
K_{{\rm Re}F}^{(2)}(\tau) =-(\kappa_{\rm out}/4)\, {\rm Re}[e^{-\tilde{\kappa}\tau/2}].
    \label{S-KF2-res}\ee
Similarly (with a little more work) we obtain
    \begin{eqnarray}
&& K_{{\rm Re}F}^{(3)}(\tau) = \frac{\kappa_{\rm out}\kappa}{4}\int_{-\infty}^0 \left[ {\rm Re}( e^{\tilde{\kappa}t'/2}) \, {\rm Re}\left( e^{-\tilde{\kappa}(\tau - t')/2}\right) \right.
    \nonumber \\
&& \hspace{2.7cm} \left.
+\,{\rm Im}( e^{\tilde{\kappa}t'/2}) \, {\rm Im}\left( e^{-\tilde{\kappa}(\tau - t')/2}\right) \right] dt'
    \nonumber \\
&& \hspace{1.5cm}
= (\kappa_{\rm out}/4) \, e^{-\kappa \tau /2} \cos(\Delta \omega \, \tau) ,
    \label{S-KF3-res}\end{eqnarray}
where addition of contributions from $v(t)$ and $v_{\rm a}(t)$ gives the coefficient $\kappa=\kappa_{\rm out}+(\kappa -\kappa_{\rm out})$ on the first line. It is easy to see that Eqs.\ (\ref{S-KF2-res}) and (\ref{S-KF3-res}) exactly cancel each other, thus proving Eq.\ (\ref{S-K-F}). The proof using the standard input-output theory \cite{S-Gardiner-book, S-Gardiner-1985, S-Clerk-2010} is essentially the same as our proof, just using operators instead of complex numbers and associating time-dependence with the Heisenberg picture.

This result explains why we do not see significant exponential contributions $\propto \exp(-\kappa_i\tau /2)$ to the self-correlators $K_{ii}(\tau )$ at small $\tau$ in Fig.\ 2(d) of the main text. However, finite bandwidth of the amplifier leads to widening of the delta-function correlator of the amplified signal, producing crudely exponential time dependence at small $\tau$ in Fig.\ 2(d) of the main text.

While we have shown above that in the ideal case the amplified noise is delta-correlated, small contributions $\propto~\exp(-\kappa_i\tau /2)$ to the self-correlators are possible due to various non-idealities. For example, if the temperature of the resonator is significant, then the delta-correlator of the noise $v_{\rm a}(t)$ is larger than the vacuum correlator of $v(t)$. Repeating the derivation above, we see that the coefficient in Eq.\ (\ref{S-KF3-res}) increases, and the cancellation by Eq.\ (\ref{S-KF2-res}) is incomplete. Similarly, if the amplitude of the microwave drive fluctuates in time, these fluctuations are essentially passed through a filter with the resonator bandwidth, creating a contribution  $\propto \exp(-\kappa_i\tau /2)$ from the ``white-noise'' part of the fluctuations.

A similar mechanism is produced by random evolution of the qubit, which is much slower than $\kappa_i$, but still has a non-zero spectral weight at frequencies comparable to $\kappa_i$. Let us estimate the corresponding contribution to the self-correlator $K_{ii}(\tau)$ in the following way. Assuming $\varphi=\pi/2$ (so that $z$ and $x$ components of the qubit are measured) and assuming $\Gamma_z=\Gamma_x=\Gamma$, let us consider the uniform diffusion of the qubit state along the $x$-$z$ great circle on the  Bloch sphere \cite{S-Ruskov2010} with the angular diffusion coefficient $2\Gamma$. Note that we assume an ideal detector by separating a non-ideal detector into an ideal part and extra noise. Even though the Markovian theory \cite{S-Ruskov2010} cannot describe the qubit evolution at the frequency scale $\kappa_i$, in this estimate we just assume the same uniform diffusion with coefficient $2\Gamma$.

In the Markovian approximation, the qubit evolution characterized by the angle $\beta(t)$ from the $x$-axis, produces the output signal in the $\sigma_z$-channel $I_z(t)=\sin\beta (t)$ (excluding noise).  However, because of the finite bandwidth $\kappa_z$ of the resonator, there will be a correction $-\int_{-\infty}^t e^{-\kappa_z(t-t')/2} \cos\beta \, \dot{\beta}\, dt' $ to the output signal due to transient delay. The contribution from this correction to the self-correlator $K_{zz}(\tau)=T^{-1}\int_{-T/2}^{T/2} I_z(t)\,I_z(t+\tau)\, dt$ (with $T\to\infty$) is $\kappa_z^{-1}e^{-\kappa_z\tau/2} \langle [ (\cos \beta) \Delta \beta ]^2/\Delta t \rangle $, where $\Delta \beta=\dot{\beta}\,\Delta t$ is the change during small $\Delta t$. Using $\langle \cos^2\beta \rangle =1/2$ and $\langle (\Delta\beta)^2\rangle =2\Gamma \Delta t$, we obtain the contribution $(\Gamma/\kappa_z)\,e^{-\kappa_z\tau/2}$ to $K_{zz}(\tau)$. A slightly more accurate calculation, which takes into account qubit diffusion during resonator transients, produces the result $(\Gamma/\kappa_z)\,e^{-\kappa_z\tau/2}e^{-\Gamma\tau}$, which is practically the same since in our case $\Gamma \ll \kappa_z$. The same crude derivation can be done for the $\sigma_x$-channel. Thus, for $\varphi=\pi/2$ we expect the contribution $\sim (\Gamma /\kappa_i)\, e^{-\kappa_i\tau/2}$ to the self-correlator $K_{ii}(\tau )$ at small $\tau$. However, these contributions are not visible in Fig.\ 2(d) of the main text because experimentally $\Gamma /\kappa_z=0.028$ and $\Gamma /\kappa_\varphi=0.017$, which is almost three orders of magnitude less than the scale of the amplifier-caused effect contributing to the lines in Fig.\ 2(d) [note that by design of the experiment $\Gamma/\kappa_i\ll 1$ -- see Eq.\ (\ref{S-ineq})].


\end{document}